\documentclass[10pt]{article}
\usepackage{graphicx}
\usepackage{amsmath}
\usepackage{amssymb}
\usepackage{caption2}
\begin{document}
\bibliographystyle{prsty}
\begin{center}
{\large {\bf \sc{  Analysis of  the hidden-charm pentaquark candidates in the $J/\psi \Omega$ mass spectrum  via the  QCD sum rules }}} \\[2mm]
Zhi-Gang Wang \footnote{E-mail: zgwang@aliyun.com.  }  \\
 Department of Physics, North China Electric Power University, Baoding 071003, P. R. China
\end{center}

\begin{abstract}
We explore the diquark-diquark-antiquark type $sssc\bar{c}$ pentaquark states with the isospin-spin-parity  $IJ^{P}=0{\frac{1}{2}}^-$, $0{\frac{3}{2}}^-$ and $0{\frac{5}{2}}^-$ via  the QCD sum rules  in details  and  obtain the ground state mass spectrum.  And we suggest to hunt for these exotic states in the exclusive  decay $\Omega_b^- \to P_{csss}^-\phi \to J/\psi \Omega^-  \phi $. Observations of the $P_{csss}$ states would shed light on the nature of  exotic states and provide an excellent opportunity to distinguish the diquark-diquark-antiquark type pentaquark states from meson-baryon type molecular states.
\end{abstract}

 PACS number: 12.39.Mk, 14.20.Lq, 12.38.Lg

Key words: Pentaquark states, QCD sum rules

\section{Introduction}
Since observation of the $X(3872)$ by the Belle collaboration in 2003 \cite{X3872-2003}, we have entered an era of burst of various hadrons including both the traditional  states and  non-traditional  exotic states, which summon a clear classification or a clear spectroscopy.
 The hadron spectroscopy  provides  us an excellent opportunity for  deep  understanding of the strong interactions in non-perturbative region, which is a challenging problem in particle physics over past decades.
As a result,  the properties of exotic states have become an attractive research field.

Especially, the striking irruptions of experimental observations/evidences of pentaquark candidates,
such as the  $P_c(4380)$ and $P_c(4450)$ in 2015 \cite{LHCb-4380},  $P_c(4312)$,  $P_c(4440)$ and $P_c(4457)$ in 2019 \cite{LHCb-Pc4312}, $P_{cs}(4459)$
in 2020 \cite{LHCb-Pcs4459-2012}, $P_c(4337)$ in 2021  \cite{LHCb-Pc4337},
$P_{cs}(4338)$ in 2022 \cite{LHCb-Pcs4338}, $P_{cs}(4459)$ in 2025 \cite{Belle-Pcs4338-Pcs4459}, have stimulated hot debates on their natures and
substructures. The theoretical physicists have suggested several feasible  interpretations,  such as diquark-diquark-antiquark or diquark-triquark type hidden-charm pentaquark states \cite{di-di-anti-penta-1,di-di-anti-penta-2,Wang1508-EPJC,
WangHuang-EPJC-1508-12,WangZG-EPJC-1509-12,WangZG-NPB-1512-32,
Pc4312-penta-3,Pc4312-penta-1,WZG-penta-IJMPA,
WangZG-Pcs4459-333,WangZG-Pc12-JpsiLambda,WangZG-Pc12-Jpsip,
WangZG-Pc12-JpsiXi,WangZG-Pc12-JpsiSgm,WangZG-Pc12-JpsiXi-10,
di-tri-penta-1,
di-tri-penta-2}, meson-baryon type molecular states \cite{mole-penta-2,mole-penta-1,mole-penta-3,mole-penta-10,
mole-penta-11,mole-penta-6,
mole-penta-5,Pc4312-mole-penta-2,
Pc4312-mole-penta-1,Pc4312-mole-penta-WXW-SCPMA,
Pc4312-mole-penta-WXW-IJMPA,Pcs4338-mole-XWWang,Pc4312-mole-penta-7,
Pcs4338-mole-FKGuo,Pcs4459-mole-FLWang,Pcs4459-mole-CWXiao,Coupled-Pc4337-MJYan-EPJC-2022,
Pcs4338-mole-LMeng}, kinematical effects \cite{ATS-LiuXH-2016,ATS-Bayar-GuoFK-2016,ATS-LiuXH-2020}. As far as the theoretical methods are concerned, there are constituent diquark models, QCD-inspired potential models,  one-boson-exchange potential models, effective Chiral Lagrangian approach,  Born-Oppenheimer approximation, (quasipotential) Bethe-Salpeter equation or Lippmann-Schwinger equation,  QCD sum rules, Lattice QCD, etc. One can consult Ref.\cite{WangZG-Review} for literatures.
 Although the hidden-charm pentaquark candidates have been explored  extensively  by  different theoretical approaches, there still exists severe ambiguity in their  properties, their precise structures remain unknown. Up to now,  there is no consensus on their natures,  and more experimental and theoretical works are still needed.

If we focus on five-quark picture, there exist a charm-anticharm quark pair $c\bar{c}$ and three light quarks $qqq$, the quantum number of strangeness might be
$S=0$, $-1$, $-2$, $-3$. Thus the $P_c(4312)$, $P_c(4337)$, $P_c(4380)$, $P_c(4440)$ and $P_c(4457)$ have strangeness $S=0$, and $P_{cs}(4338)$ and
$ P_{cs}(4459)$ have strangeness $S=-1$, but there exists no experimental candidate for the hidden-charm pentaquark states with strangeness $S=-2$, $-3$. On the theoretical side, the hadron spectroscopy with strangeness $S=-3$ is rarely investigated  so far \cite{WangZG-EPJC-1509-12,WangZG-NPB-1512-32,Zong-penta-sss-PLB-19,Li-penta-sss-PRD-23,WangFL-mole-sss-PRD-21,
Clymton-mole-sss-PRD-25,Oset-mole-sss-PRD-24,Azizi-penta-sss-PRD-23}.

Still further, we restrict our interest on the diquark-diquark-antiquark type  hidden-charm pentaquark states, where all the pentaquark candidates, such as the $P_c(4312)$, $P_c(4337)$, $P_{cs}(4338)$, $P_{c}(4380)$, $P_c(4440)$, $P_c(4457)$, $P_{cs}(4459)$, could find their suitable positions to sit down \cite{WangZG-Pc12-JpsiLambda,WangZG-Pc12-Jpsip}.
It is natural and interesting to
hunt for the hidden-charm pentaquark candidates with  strangeness $S=-1$ in the $J/\psi\Lambda$, $J/\psi\Sigma$, $J/\psi\Sigma^*$ invariant mass distributions, $S=-2$ in the $J/\psi\Xi$ and $J/\psi \Xi^*$ invariant mass distributions or $S=-3$ in the $J/\psi\Omega$ invariant mass distribution \cite{di-di-anti-penta-1,di-di-anti-penta-2,
WangZG-EPJC-1509-12,WangZG-NPB-1512-32,Pc4312-penta-1}.

The QCD sum rules method is one of the most powerful theoretical tool in studying the exotic states \cite{WangZG-Review,WangZG-landau-PRD}, such as the hidden-charm tetraquark states, pentaquark states, molecular states, etc.
 After  observation  of the $P_c(4380)$ and $P_c(4450)$,
  we  studied  the diquark-diquark-antiquark type  hidden-charm pentaquark states with the spin-parity  $J^P={\frac{1}{2}}^\pm$, ${\frac{3}{2}}^\pm$, ${\frac{5}{2}}^\pm$  and strangeness   $S=0,\,-1,\,-2,\,-3$  by carrying out the operator product expansion (OPE) up to the vacuum condensates of dimension 10 and neglecting small contributions of the gluon condensates  \cite{Wang1508-EPJC,WangHuang-EPJC-1508-12,WangZG-EPJC-1509-12,
 WangZG-NPB-1512-32}. These works are valuable at least as rough estimations of the hidden-charm pentaquark spectroscopy.

After observations/evidences  of the $P_c(4312)$, $P_{cs}(4338)$ and   $P_{cs}(4459)$,   we updated our  old calculations  \cite{Wang1508-EPJC,WangHuang-EPJC-1508-12,WangZG-EPJC-1509-12,
 WangZG-NPB-1512-32} by carrying out the OPE up to  the   vacuum condensates of dimension $13$  consistently,  and  exhausted the ground state spectroscopy of  the $uudc\bar{c}$, $udsc\bar{c}$, $qssc\bar{c}$ and $qqsc\bar{c}$ pentaquark states  with the spin-parity $J^P={\frac{1}{2}}^-$, ${\frac{3}{2}}^-$, ${\frac{5}{2}}^-$ and isospins $I=\frac{1}{2}$, $0$, $\frac{1}{2}$ and $1$ respectively, made reasonable  identifications of the $P_c(4312)$, $P_c(4337)$, $P_{cs}(4338)$, $P_c(4380)$, $P_c(4440)$, $P_c(4457)$  and
$ P_{cs}(4459)$, and suggested to hunt for these exotic pentaquark states in the $J/\psi p$, $J/\psi \Delta$, $J/\psi\Lambda$, $J/\psi\Xi$, $J/\psi\Sigma$, $J/\psi\Xi^*$ invariant mass distributions \cite{WZG-penta-IJMPA,WangZG-Pc12-JpsiLambda,WangZG-Pc12-Jpsip,
WangZG-Pc12-JpsiXi,WangZG-Pc12-JpsiSgm,WangZG-Pc12-JpsiXi-10}.

 Now  we extend our previous works to explore the lowest  diquark-diquark-antiquark type $sssc\bar{c}$ pentaquark states with the isospin-spin-parity $IJ^P=0{\frac{1}{2}}^-$, $0{\frac{3}{2}}^-$ and $0{\frac{5}{2}}^-$ in details  via the QCD sum rules, and suggest to hunt for these hidden-charm pentaquark states with strangeness $S=-3$ in the $J/\psi\Omega$ invariant mass distribution, their observations   are of crucial importance in diagnosing  the exotic states
and in distinguishing the compact pentaquark states and loose molecular states.

 The article is arranged as follows:  we obtain the QCD sum rules for the  $sssc\bar{c}$ pentaquark states in Sect.2;  in Sect.3, we present the numerical results and discussions; and Sect.4 is reserved for our
conclusion.

\section{QCD sum rules for  the  $sssc\bar{c}$ pentaquark states}
Firstly, let us write down  the two-point correlation functions $\Pi(p)$, $\Pi_{\mu\nu}(p)$ and $\Pi_{\mu\nu\alpha\beta}(p)$ routinely,
\begin{eqnarray}\label{CF-Pi-Pi-Pi}
\Pi(p)&=&i\int d^4x e^{ip \cdot x} \langle0|T\left\{J(x)\bar{J}(0)\right\}|0\rangle \, ,\nonumber\\
\Pi_{\mu\nu}(p)&=&i\int d^4x e^{ip \cdot x} \langle0|T\left\{J_{\mu}(x)\bar{J}_{\nu}(0)\right\}|0\rangle \, ,\nonumber\\
\Pi_{\mu\nu\alpha\beta}(p)&=&i\int d^4x e^{ip \cdot x} \langle0|T\left\{J_{\mu\nu}(x)\bar{J}_{\alpha\beta}(0)\right\}|0\rangle \, ,
\end{eqnarray}
where
 \begin{eqnarray}
 J(x)&=&J^1(x)\, , \, J^2(x)\, , \nonumber\\
 J_\mu(x)&=&J_\mu^1(x)\, , \, J_\mu^2(x)\, , \, J_\mu^3(x)\, ,  \nonumber\\
 J_{\mu\nu}(x)&=&J_{\mu\nu}^1(x)\, , \, J_{\mu\nu}^2(x)\, ,
 \end{eqnarray}
 \begin{eqnarray}\label{Current-12}
J^{1}(x)&=&\varepsilon^{ila} \varepsilon^{ijk}\varepsilon^{lmn} s^T_j(x) C\gamma_\mu s_k(x)s^T_m(x) C\gamma^\mu c_n(x)  C\bar{c}^{T}_{a}(x) \, , \nonumber\\
J^{2}(x)&=&\varepsilon^{ila} \varepsilon^{ijk}\varepsilon^{lmn} s^T_j(x) C\gamma_\mu s_k(x) s^T_m(x) C\gamma_5 c_n(x) \gamma_5 \gamma^\mu  C\bar{c}^{T}_{a}(x) \, ,
 \end{eqnarray}
with the isospin-spin $(I,J)=(0,\frac{1}{2})$,
\begin{eqnarray}\label{Current-32}
 J^{1}_{\mu}(x)&=&\varepsilon^{ila} \varepsilon^{ijk}\varepsilon^{lmn} s^T_j(x) C\gamma_\mu s_k(x) s^T_m(x) C\gamma_5 c_n(x)    C\bar{c}^{T}_{a}(x) \, , \nonumber \\
 J^{2}_{\mu}(x)&=&\varepsilon^{ila} \varepsilon^{ijk}\varepsilon^{lmn} s^T_j(x) C\gamma_\mu s_k(x)s^T_m(x) C\gamma_\alpha c_n(x)\gamma_5\gamma^\alpha C\bar{c}^{T}_{a}(x) \, , \nonumber\\
J^{3}_{\mu}(x)&=&\varepsilon^{ila} \varepsilon^{ijk}\varepsilon^{lmn} s^T_j(x) C\gamma_\alpha s_k(x)s^T_m(x) C\gamma_\mu c_n(x) \gamma_5\gamma^\alpha C\bar{c}^{T}_{a}(x) \, ,
 \end{eqnarray}
 with the isospin-spin $(I,J)=(0,\frac{3}{2})$,
\begin{eqnarray} \label{Current-52}
J^1_{\mu\nu}(x)&=&\frac{\varepsilon^{ila} \varepsilon^{ijk}\varepsilon^{lmn} }{\sqrt{2}}\, s^T_j(x) C\gamma_\mu s_k(x)\, s^T_m(x) C\gamma_5 c_n(x)\, \gamma_5\gamma_{\nu}C\bar{c}^{T}_{a}(x)+(\mu \leftrightarrow \nu) \, , \nonumber\\
J^2_{\mu\nu}(x)&=&\frac{\varepsilon^{ila} \varepsilon^{ijk}\varepsilon^{lmn}}{\sqrt{2}} s^T_j(x) C\gamma_\mu s_k(x)\, s^T_m(x) C\gamma_\nu c_n(x)  C\bar{c}^{T}_{a}(x)+(\mu \leftrightarrow \nu)\, ,
\end{eqnarray}
with the isospin-spin $(I,J)=(0,\frac{5}{2})$,
 the $i$, $j$, $k$, $l$, $m$, $n$ and $a$ are color indexes. There are three strange quarks, it is not necessary to mention the isospin, but to be consistent with our previous works \cite{WangZG-Pc12-JpsiLambda,WangZG-Pc12-Jpsip,
WangZG-Pc12-JpsiXi,WangZG-Pc12-JpsiSgm,WangZG-Pc12-JpsiXi-10,Pc4312-mole-penta-WXW-SCPMA,
Pc4312-mole-penta-WXW-IJMPA,Pcs4338-mole-XWWang}, where the isospin is emphasized, we retain this somewhat redundant quantum number.
The hidden-charm pentaquark states have three light valence quarks,   the $qqq$ components have the light-flavor $SU(3)$ symmetry,
\begin{eqnarray}\label{two-octet}
{\mathbf{3}}\otimes {\mathbf{3}}\otimes {\mathbf{3}} &\to & \left({\bar{\mathbf{3}}} \oplus {\mathbf{6}} \right) \otimes {\mathbf{3}}\, , \nonumber\\
&\to& {\mathbf{1}} \oplus {\mathbf{8}}_1 \oplus {\mathbf{8}}_2\oplus {\mathbf{10}}\, ,
\end{eqnarray}
there exist one singlet, two octets and one decuplet hidden-charm pentaquark states. It is obvious that the three $s$-quarks in the currents $J(x)$, $J_\mu(x)$ and $J_{\mu\nu}(x)$ have to be symmetric to warrant that  the currents $J(x)$, $J_\mu(x)$ and $J_{\mu\nu}(x)$ are in the light-flavor $\mathbf{10}$ representation.

The light (L)  diquark $\varepsilon^{ijk}s^T_jC\gamma_{\mu}s_k$ has spin $S_L=1$, while the heavy (H) diquarks  $\varepsilon^{ijk}s^T_jC\gamma_5c_k$ and   $\varepsilon^{ijk}s^T_jC\gamma_{\mu}c_k$ have spins $S_H=0$ and $1$, respectively. A heavy and  a light  diquark attract each other and a tetraquark in the color triplet comes into being,  the  angular momentum $\vec{J}_{LH}=\vec{S}_L+\vec{S}_H$ and $J_{LH}=0$, $1$, $2$.
The anti-charm quark operators $C\bar{c}_a^T$ and $\gamma_5\gamma_{\mu}C\bar{c}_a^T$ have the spin-parity $J^P={\frac{1}{2}}^-$ and ${\frac{3}{2}}^-$, respectively. The total angular momentums   $\vec{J}=\vec{J}_{LH}+\vec{J}_{\bar{c}}$ and $J=\frac{1}{2}$, $\frac{3}{2}$, $\frac{5}{2}$, which are illustrated   clearly  in Table \ref{current-pentaQ}.

\begin{table}
\begin{center}
\begin{tabular}{|c|c|c|c|c|c|c|c|c|}\hline\hline
$[qq][qc]\bar{c}$ ($S_L$, $S_H$, $J_{LH}$, $J$)  & $J^{P}$             & Currents              \\ \hline

$[ss][sc]\bar{c}$ ($1$, $1$, $0$, $\frac{1}{2}$) &${\frac{1}{2}}^{-}$  &$J^1(x)$        \\

$[ss][sc]\bar{c}$ ($1$, $0$, $1$, $\frac{1}{2}$) &${\frac{1}{2}}^{-}$  &$J^2(x)$         \\    \hline

$[ss][sc]\bar{c}$ ($1$, $0$, $1$, $\frac{3}{2}$) &${\frac{3}{2}}^{-}$ &$J^1_\mu(x)$          \\

$[ss][sc]\bar{c}$ ($1$, $1$, $2$, $\frac{3}{2}$)${}_2$ &${\frac{3}{2}}^{-}$  &$J^2_\mu(x)$   \\

$[ss][sc]\bar{c}$ ($1$, $1$, $2$, $\frac{3}{2}$)${}_3$ &${\frac{3}{2}}^{-}$  &$J^3_\mu(x)$   \\ \hline

$[ss][sc]\bar{c}$ ($1$, $0$, $1$, $\frac{5}{2}$) &${\frac{5}{2}}^{-}$  &$J^1_{\mu\nu}(x)$    \\

$[ss][sc]\bar{c}$ ($1$, $1$, $2$, $\frac{5}{2}$) &${\frac{5}{2}}^{-}$  &$J^2_{\mu\nu}(x)$   \\
\hline\hline
\end{tabular}
\end{center}
\caption{ The valence quark structures and quantum numbers of the currents.  }\label{current-pentaQ}
\end{table}

The currents $J(x)$, $J_\mu(x)$ and $J_{\mu\nu}(x)$ have the spin-parity
$J^P={\frac{1}{2}}^-$, ${\frac{3}{2}}^-$ and ${\frac{5}{2}}^-$, respectively, and
 couple potentially to the $sssc\bar{c}$  pentaquark states (P) having both negative  and positive parities on the basis of quark-hadron duality  \cite{Wang1508-EPJC,WangZG-Review},
\begin{eqnarray}\label{Coupling12}
\langle 0| J (0)|P_{\frac{1}{2}}^{-}(p)\rangle &=&\lambda^{-}_{\frac{1}{2}} U^{-}(p,s) \, , \nonumber \\
\langle 0| J (0)|P_{\frac{1}{2}}^{+}(p)\rangle &=&\lambda^{+}_{\frac{1}{2}} i\gamma_5 U^{+}(p,s) \, ,
\end{eqnarray}
\begin{eqnarray}
\langle 0| J_{\mu} (0)|P_{\frac{3}{2}}^{-}(p)\rangle &=&\lambda^{-}_{\frac{3}{2}} U^{-}_\mu(p,s) \, ,  \nonumber \\
\langle 0| J_{\mu} (0)|P_{\frac{3}{2}}^{+}(p)\rangle &=&\lambda^{+}_{\frac{3}{2}}i\gamma_5 U^{+}_\mu(p,s) \, ,
\end{eqnarray}
\begin{eqnarray}\label{Coupling52}
\langle 0| J_{\mu\nu} (0)|P_{\frac{5}{2}}^{-}(p)\rangle &=&\sqrt{2}\lambda^{-}_{\frac{5}{2}} U^{-}_{\mu\nu}(p,s) \, ,\nonumber\\
\langle 0| J_{\mu\nu} (0)|P_{\frac{5}{2}}^{+}(p)\rangle &=&\sqrt{2}\lambda^{+}_{\frac{5}{2}}i\gamma_5 U^{+}_{\mu\nu}(p,s) \, ,
\end{eqnarray}
where the superscripts $\pm$ stand for  the  parities, the subscripts $\frac{1}{2}$, $\frac{3}{2}$ and $\frac{5}{2}$ stand for  the spins,     the $\lambda$ are the pole residues. The $U^\pm(p,s)$,  $U^{\pm}_\mu(p,s)$ and $U^{\pm}_{\mu\nu}(p,s)$ are Dirac and Rarita-Schwinger spinors respectively \cite{Wang1508-EPJC,WangZG-Review}.

Routinely, we insert  a complete set  of intermediate
$sssc\bar{c}$ pentaquark states with the same quantum numbers as the currents  $J(x)$, $i\gamma_5 J(x)$, $J_{\mu}(x)$, $i\gamma_5 J_{\mu}(x)$, $J_{\mu\nu}(x)$  and $i\gamma_5 J_{\mu\nu}(x)$ into the correlation functions to achieve the hadronic representation through dispersion relation
\cite{SVZ79-1,SVZ79-2,PRT85},  isolate the  lowest  states owning to negative and positive parities (for detailed derivations, one can consult Ref.\cite{WangZG-Pc12-JpsiSgm}),
\begin{eqnarray}\label{CF-Hadron-12}
\Pi(p) & = & {\lambda^{-}_{\frac{1}{2}}}^2  {\!\not\!{p}+ M_{-} \over M_{-}^{2}-p^{2}  }+  {\lambda^{+}_{\frac{1}{2}}}^2  {\!\not\!{p}- M_{+} \over M_{+}^{2}-p^{2}  } +\cdots  \, ,\nonumber\\
&=&\Pi_{\frac{1}{2}}^1(p^2)\!\not\!{p}+\Pi_{\frac{1}{2}}^0(p^2)\, ,
 \end{eqnarray}
\begin{eqnarray}\label{CF-Hadron-32}
 \Pi_{\mu\nu}(p) & = & {\lambda^{-}_{\frac{3}{2}}}^2  {\!\not\!{p}+ M_{-} \over M_{-}^{2}-p^{2}  } \left(- g_{\mu\nu}
\right)+  {\lambda^{+}_{\frac{3}{2}}}^2  {\!\not\!{p}- M_{+} \over M_{+}^{2}-p^{2}  } \left(- g_{\mu\nu}\right)  +\cdots \nonumber \\
&=&\left[\Pi_{\frac{3}{2}}^1(p^2)\!\not\!{p}+\Pi_{\frac{3}{2}}^0(p^2)\right]\left(- g_{\mu\nu}\right)+\cdots\, ,
\end{eqnarray}
\begin{eqnarray}\label{CF-Hadron-52}
\Pi_{\mu\nu\alpha\beta}(p) & = &2{\lambda^{-}_{\frac{5}{2}}}^2  {\!\not\!{p}+ M_{-} \over M_{-}^{2}-p^{2}  } \frac{ g_{\mu\alpha}g_{\nu\beta}
+g_{\mu\beta}g_{\nu\alpha}}{2}
+  2 {\lambda^{+}_{\frac{5}{2}}}^2  {\!\not\!{p}- M_{+} \over M_{+}^{2}-p^{2}  } \frac{ g_{\mu\alpha}g_{\nu\beta}
+g_{\mu\beta}g_{\nu\alpha}}{2}
     +\cdots \, , \nonumber\\
& = & \left[\Pi_{\frac{5}{2}}^1(p^2)\!\not\!{p}+\Pi_{\frac{5}{2}}^0(p^2)\right]\left( g_{\mu\alpha}g_{\nu\beta}+g_{\mu\beta}g_{\nu\alpha}\right)  +\cdots \, .
 \end{eqnarray}
 We retain  components $\Pi_{\frac{1}{2}}^{1/0}(p^2)$, $\Pi_{\frac{3}{2}}^{1/0}(p^2)$  and  $\Pi_{\frac{5}{2}}^{1/0}(p^2)$ to study the $sssc\bar{c}$ pentaquark states with the spins $J=\frac{1}{2}$, $\frac{3}{2}$ and $\frac{5}{2}$, respectively.

Then we obtain the hadronic (H) spectral densities  through  dispersion relation,
\begin{eqnarray}
\frac{{\rm Im}\Pi^1_J(s)}{\pi}&=& \lambda_{-}^2 \delta\left(s-M_{-}^2\right)+\lambda_{+}^2 \delta\left(s-M_{+}^2\right) =\, \rho^1_{H}(s) \, , \\
\frac{{\rm Im}\Pi^0_J(s)}{\pi}&=&M_{-}\lambda_{-}^2 \delta\left(s-M_{-}^2\right)-M_{+}\lambda_{+}^2 \delta\left(s-M_{+}^2\right)
=\rho^0_{H}(s) \, ,
\end{eqnarray}
with the spins $J=\frac{1}{2}$, $\frac{3}{2}$, $\frac{5}{2}$.
We multiply  them with   weight functions $\sqrt{s}\exp\left(-\frac{s}{T^2}\right)$ and $\exp\left(-\frac{s}{T^2}\right)$ to get the QCD sum rules,
\begin{eqnarray}
\int_{4m_c^2}^{s_0}ds \left[\sqrt{s}\,\rho^1_{H}(s)+\rho^0_{H}(s)\right]\exp\left( -\frac{s}{T^2}\right)
&=&2M_{-}\lambda_{-}^2\exp\left( -\frac{M_{-}^2}{T^2}\right) \, ,
\end{eqnarray}
\begin{eqnarray}
\int_{4m_c^2}^{s^\prime_0}ds \left[\sqrt{s}\,\rho^1_{H}(s)-\rho^0_{H}(s)\right]\exp\left( -\frac{s}{T^2}\right)
&=&2M_{+}\lambda_{+}^2\exp\left( -\frac{M_{+}^2}{T^2}\right) \, ,
\end{eqnarray}
where the $s_0$ and $s_0^\prime$ are the continuum threshold parameters,  and the $T^2$ is the Borel parameter. We  separate  contributions  of the $sssc\bar{c}$ pentaquark states owning to negative and positive parities clearly \cite{WangZG-Review}.

On the QCD side,  we carry out  the OPE up to the quark-gluon operators of dimension $13$ and order $\mathcal{O}( \alpha_s^{k})$ with $k\leq 1$  consistently to get the vacuum condensates through vacuum saturation, and   take the terms  $\propto m_s$ to symbolize   the light-flavor  $SU(3)$ mass-breaking effects \cite{WangZG-Review}.
 Again, we get  the spectral densities through   dispersion relation routinely,
\begin{eqnarray}\label{QCD-rho}
 \rho^1_{QCD}(s) &=&\frac{{\rm Im}\Pi^1_J(s)}{\pi}\, , \nonumber\\
\rho^0_{QCD}(s) &=&\frac{{\rm Im}\Pi^0_J(s)}{\pi}\, ,
\end{eqnarray}
with the spins $J=\frac{1}{2}$, $\frac{3}{2}$, $\frac{5}{2}$.
As an example, we present  explicit expressions of the QCD spectral densities for the current $J^1(x)$ in  Appendix,  and neglect  length expressions of other spectral densities for simplicity.

Finally,  we  match the hadron side with the QCD side of the  spectral representations, complete  quark-hadron duality below the continuum thresholds, and  achieve  two  QCD sum rules:
\begin{eqnarray}\label{QCDSR}
2M_{-}\lambda_{-}^2\exp\left( -\frac{M_{-}^2}{T^2}\right)&=& \int_{4m_c^2}^{s_0}ds \,\left[\sqrt{s}\rho_{QCD}^1(s)+\rho_{QCD}^{0}(s)\right]\,\exp\left( -\frac{s}{T^2}\right)\,  ,
\end{eqnarray}
\begin{eqnarray}\label{QCDSR-Positive}
2M_{+}\lambda_{+}^2\exp\left( -\frac{M_{+}^2}{T^2}\right)&=& \int_{4m_c^2}^{s^\prime_0}ds \,\left[\sqrt{s}\rho_{QCD}^1(s)-\rho_{QCD}^{0}(s)\right]\,\exp\left( -\frac{s}{T^2}\right)\,  .
\end{eqnarray}

We adopt the QCD sum rules for the $sssc\bar{c}$ pentaquark states  with negative parity, and differentiate  Eq.\eqref{QCDSR} in regard   to  $\frac{1}{T^2}$ to get  the explicit formula for the hadron  masses,
 \begin{eqnarray}\label{QCDSR-mass}
 M^2_{-} &=& \frac{-\int_{4m_c^2}^{s_0}ds \frac{d}{d(1/T^2)}\, \left[\sqrt{s}\rho_{QCD}^1(s)+\rho_{QCD}^{0}(s)\right]\,\exp\left( -\frac{s}{T^2}\right)}{\int_{4m_c^2}^{s_0}ds \, \left[\sqrt{s}\rho_{QCD}^1(s)+\rho_{QCD}^{0}(s)\right]\,\exp\left( -\frac{s}{T^2}\right)}\,  ,
\end{eqnarray}
thereafter, we would like to use $M_P$ ($\lambda_P$) in stead of $M_{-}$ ($\lambda_{-}$) to denote the masses (pole residues) of the hidden-charm pentaquark states with the negative parity.

\section{Numerical results and discussions}
At the beginning points, we take  the traditional  values of the  vacuum condensates
$\langle\bar{q}q \rangle=-(0.24\pm 0.01\, \rm{GeV})^3$,  $\langle\bar{s}s \rangle=(0.8\pm0.1)\langle\bar{q}q \rangle$,
 $\langle\bar{s}g_s\sigma G s \rangle=m_0^2\langle \bar{s}s \rangle$,
$m_0^2=(0.8 \pm 0.1)\,\rm{GeV}^2$, $\langle \frac{\alpha_s
GG}{\pi}\rangle=0.012\pm0.004\,\rm{GeV}^4$    at the special energy scale  $\mu=1\, \rm{GeV}$
\cite{SVZ79-1,SVZ79-2,PRT85,ColangeloReview}, and  take the $\overline{MS}$ (modified-minimal-subtraction) quark  masses $m_{c}(m_c)=(1.275\pm0.025)\,\rm{GeV}$
 and $m_s(\mu=2\,\rm{GeV})=(0.095\pm0.005)\,\rm{GeV}$
 listed in The Review of Particle Physics \cite{PDG}.
Furthermore,  we consider  their energy-scale dependence from the re-normalization group equation  \cite{Narison-mix},
 \begin{eqnarray}
  \langle\bar{s}s \rangle(\mu)&=&\langle\bar{s}s \rangle({\rm 1 GeV})\left[\frac{\alpha_{s}({\rm 1 GeV})}{\alpha_{s}(\mu)}\right]^{\frac{12}{33-2n_f}}\, , \nonumber\\
   \langle\bar{s}g_s \sigma Gs \rangle(\mu)&=&\langle\bar{s}g_s \sigma Gs \rangle({\rm 1 GeV})\left[\frac{\alpha_{s}({\rm 1 GeV})}{\alpha_{s}(\mu)}\right]^{\frac{2}{33-2n_f}}\, ,\nonumber\\
m_c(\mu)&=&m_c(m_c)\left[\frac{\alpha_{s}(\mu)}{\alpha_{s}(m_c)}\right]^{\frac{12}{33-2n_f}} \, ,\nonumber\\
m_s(\mu)&=&m_s({\rm 2GeV} )\left[\frac{\alpha_{s}(\mu)}{\alpha_{s}({\rm 2GeV})}\right]^{\frac{12}{33-2n_f}}\, ,\nonumber\\
\alpha_s(\mu)&=&\frac{1}{b_0t}\left[1-\frac{b_1}{b_0^2}\frac{\log t}{t} +\frac{b_1^2(\log^2{t}-\log{t}-1)+b_0b_2}{b_0^4t^2}\right]\, ,
\end{eqnarray}
with $t=\log \frac{\mu^2}{\Lambda^2}$, $b_0=\frac{33-2n_f}{12\pi}$, $b_1=\frac{153-19n_f}{24\pi^2}$, $b_2=\frac{2857-\frac{5033}{9}n_f+\frac{325}{27}n_f^2}{128\pi^3}$,  $\Lambda_{QCD}=210\,\rm{MeV}$, $292\,\rm{MeV}$  and  $332\,\rm{MeV}$ for the flavors  $n_f=5$, $4$ and $3$, respectively  \cite{PDG}.

We take the  flavor numbers $n_f=4$, then evolve all the  parameters to a pertinent  energy scale $\mu$, which meets  with  the modified energy scale formula,
\begin{eqnarray}
\mu &=&\sqrt{M_{P}^2-(2{\mathbb{M}}_c)^2}-3{\mathbb{M}}_s \, ,
 \end{eqnarray}
 with the effective quark masses ${\mathbb{M}}_c$ and ${\mathbb{M}}_s$. The  ${\mathbb{M}}_c$ and ${\mathbb{M}}_s$ characterize  the heavy flavor degrees of freedom and light-flavor $SU(3)$ breaking effects, respectively,  the  optimal  values fitted by previous  QCD sum rules are ${\mathbb{M}}_c=1.82\,\rm{GeV}$ and ${\mathbb{M}}_s=0.15\,\rm{GeV}$, respectively \cite{WangZG-Pcs4459-333,WangZG-Pc12-JpsiLambda,Pcs4338-mole-XWWang,
 WangZG-Review,Wang-tetra-NPB-HCss,WangZG-IJMPA-2021,
 Wang-tetra-formula,WangZG-mole-formula-1,WangZG-mole-formula-2}.

We  constraint the continuum threshold parameters as $\sqrt{s_0}=M_{P}+ (0.5-0.8)\,\rm{GeV}$ as usual \cite{WZG-penta-IJMPA,WangZG-Pc12-JpsiLambda,WangZG-Pc12-Jpsip,
WangZG-Pc12-JpsiXi,WangZG-Pc12-JpsiSgm,WangZG-Pc12-JpsiXi-10}, then
 get the  Borel  windows and continuum threshold parameters through  multitudinous  trial  and error, and present them clearly in Table \ref{Borel}. The pole contributions are about $(40-60)\%$, while the central values are slightly larger than $50\%$, and  the pole contributions are defined routinely,
\begin{eqnarray}
{\rm{pole}}&=&\frac{\int_{4m_{c}^{2}}^{s_{0}}ds\,\rho_{QCD}\left(s\right)\exp\left(-\frac{s}{T^{2}}\right)} {\int_{4m_{c}^{2}}^{\infty}ds\,\rho_{QCD}\left(s\right)\exp\left(-\frac{s}{T^{2}}\right)}\, ,
\end{eqnarray}
 with $\rho_{QCD}=\sqrt{s}\rho_{QCD}^1(s)+\rho_{QCD}^{0}(s)$.

\begin{figure}
\centering
\includegraphics[totalheight=9cm,width=11cm]{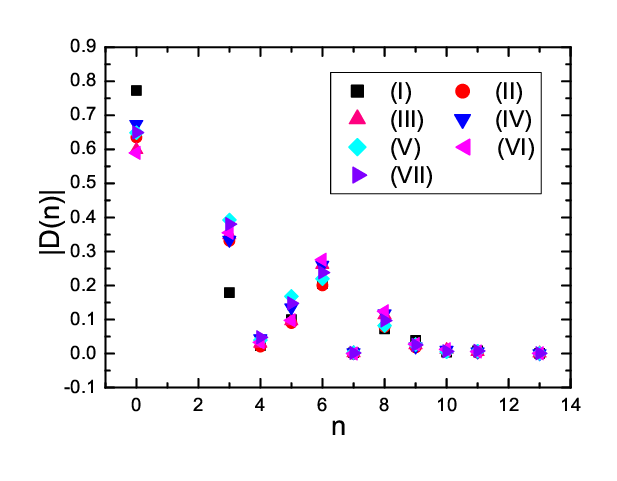}
  \caption{ The $|D(n)|$ according to variations of dimensions  $n$ with  central values of the input parameters, where the (I), (II), (III), (IV), (V), (VI)  and (VII)  denote the    $[ss][sc]\bar{c}$ ($1$, $1$, $0$, $\frac{1}{2}$),
$[ss][sc]\bar{c}$ ($1$, $0$, $1$, $\frac{1}{2}$),
$[ss][sc]\bar{c}$ ($1$, $0$, $1$, $\frac{3}{2}$),
$[ss][sc]\bar{c}$ ($1$, $1$, $2$, $\frac{3}{2}$)${}_2$,
$[ss][sc]\bar{c}$ ($1$, $1$, $2$, $\frac{3}{2}$)${}_3$,
$[ss][sc]\bar{c}$ ($1$, $0$, $1$, $\frac{5}{2}$) and
$[ss][sc]\bar{c}$ ($1$, $1$, $2$, $\frac{5}{2}$)  pentaquark states, respectively. }\label{OPE-fig}
\end{figure}

 In Fig.\ref{OPE-fig}, we plot  absolute values of contributions of the vacuum condensates via variations of their dimensions $n$ under the condition of  central values of  other  parameters, and  the $D(n)$ are defined routinely,
   \begin{eqnarray}
D(n)&=&\frac{\int_{4m_{c}^{2}}^{s_{0}}ds\,\rho_{QCD,n}(s)\exp\left(-\frac{s}{T^{2}}\right)}
{\int_{4m_{c}^{2}}^{s_{0}}ds\,\rho_{QCD}\left(s\right)\exp\left(-\frac{s}{T^{2}}\right)}\, .
\end{eqnarray}
The largest contributions are $D(0)$, $D(3)$ and $D(6)$, and they have the hierarchy $D(0)\gg D(3)>D(6)$ obviously,   the $D(4)$ and $D(7)$ are small enough due to the small value of the gluon condensate and even could be neglected safely,  the $D(6)$  serves as a milestone and characterizes the convergent behaviors of the OPE. The absolute contributions $|D(n)|$  with $n\geq 6$ have the hierarchies,
\begin{eqnarray}
&&D(6)\gg |D(8)| \gg D(9) \gg D(10)\sim|D(11)| \gg D(13) \, ,
\end{eqnarray}
  the OPE converges  very well.

\begin{table}
\begin{center}
\begin{tabular}{|c|c|c|c|c|c|c|c|}\hline\hline
                  &$T^2(\rm{GeV}^2)$     &$\sqrt{s_0}(\rm{GeV})$    &$\mu(\rm{GeV})$  &pole          &$D(13)$         \\ \hline

$J^1(x)$          &$3.8-4.2$             &$5.60\pm0.10$             &$2.8$            &$(41-60)\%$   &$\ll 1\%$      \\ \hline

$J^2(x)$          &$4.1-4.5$             &$5.69\pm0.10$             &$2.9$            &$(42-60)\%$   &$\ll 1\%$       \\ \hline

$J^1_\mu(x)$      &$4.0-4.4$             &$5.62\pm0.10$             &$2.8$            &$(43-61)\%$   &$\ll 1\%$     \\ \hline

$J^2_\mu(x)$      &$4.1-4.5$             &$5.72\pm0.10$             &$3.0$            &$(43-61)\%$   &$\ll1\%$     \\ \hline

$J^3_\mu(x)$      &$3.9-4.3$             &$5.63\pm0.10$             &$2.9$            &$(43-62)\%$   &$\ll1\%$     \\ \hline

$J^1_{\mu\nu}(x)$ &$3.9-4.3$             &$5.59\pm0.10$             &$2.8$            &$(44-62)\%$   &$\ll1\%$     \\ \hline

$J^2_{\mu\nu}(x)$ &$4.1-4.5$             &$5.70\pm0.10$             &$2.9$            &$(43-61)\%$   &$\ll1\%$     \\ \hline
\hline
\end{tabular}
\end{center}
\caption{ The Borel  windows, continuum threshold parameters, pertinent  energy scales, pole contributions and   contributions of the $D(13)$  for the $sssc\bar{c}$ pentaquark states with negative parity. }\label{Borel}
\end{table}

\begin{table}
\begin{center}
\begin{tabular}{|c|c|c|c|c|c|c|c|c|}\hline\hline
$[qq][qc]\bar{c}$ ($S_L$, $S_H$, $J_{LH}$, $J$) &$M(\rm{GeV})$   &$\lambda(10^{-3}\rm{GeV}^6)$ &Refs.\cite{WangZG-EPJC-1509-12,WangZG-NPB-1512-32}        \\ \hline

$[ss][sc]\bar{c}$ ($1$, $1$, $0$, $\frac{1}{2}$)  &$4.90\pm0.11$ &$8.60\pm1.38$                &$4.68 \pm 0.13$\\ \hline

$[ss][sc]\bar{c}$ ($1$, $0$, $0$, $\frac{1}{2}$)  &$4.97\pm0.10$ &$9.91\pm1.43$                &$4.71 \pm 0.11$\\ \hline

$[ss][sc]\bar{c}$ ($1$, $0$, $1$, $\frac{3}{2}$)  &$4.90\pm0.10$ &$4.83\pm0.70$                &$4.70 \pm 0.11$\\ \hline

$[ss][sc]\bar{c}$ ($1$, $1$, $2$, $\frac{3}{2}$)${}_2$ &$5.00\pm0.10$  &$9.71\pm1.37$ &$4.71 \pm 0.11$  \\ \hline

$[ss][sc]\bar{c}$ ($1$, $1$, $2$, $\frac{3}{2}$)${}_3$ &$4.92\pm0.11$   &$8.57\pm1.26$ &$4.72 \pm 0.11 $   \\ \hline

$[ss][sc]\bar{c}$ ($1$, $0$, $1$, $\frac{5}{2}$)  &$4.87\pm0.11$ &$4.61\pm0.67$                     &\\ \hline

$[ss][sc]\bar{c}$ ($1$, $1$, $2$, $\frac{5}{2}$)  &$4.98\pm0.10$ &$5.12\pm0.72$                  &\\ \hline\hline
\end{tabular}
\end{center}
\caption{ The masses  and pole residues of the $sssc\bar{c}$ pentaquark states, where we present our old calculations, the unit of the masses is GeV. }\label{mass-Pcs}
\end{table}

\begin{figure}
\centering
\includegraphics[totalheight=5cm,width=7cm]{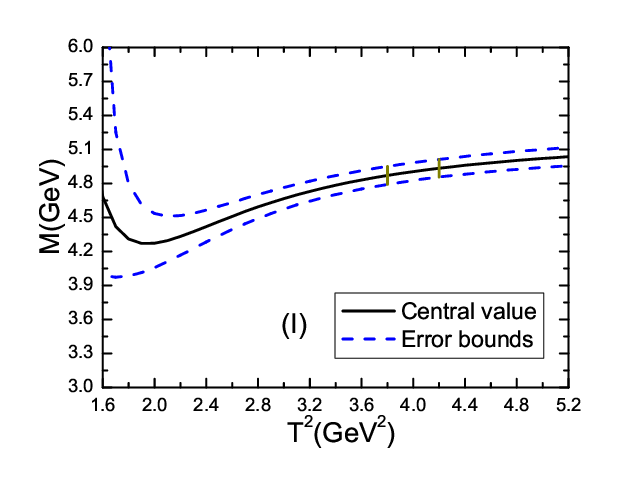}
\includegraphics[totalheight=5cm,width=7cm]{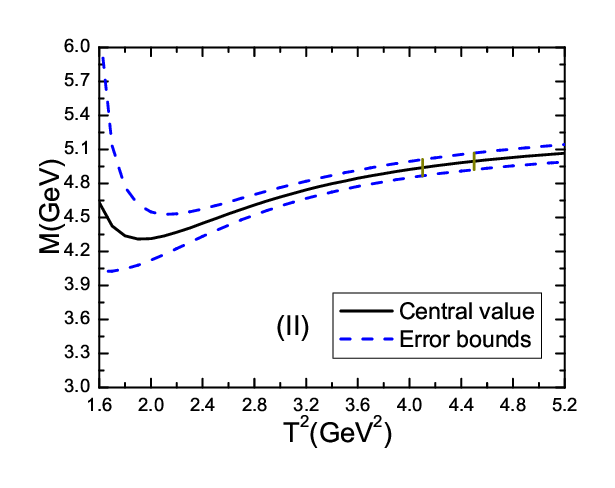}
\includegraphics[totalheight=5cm,width=7cm]{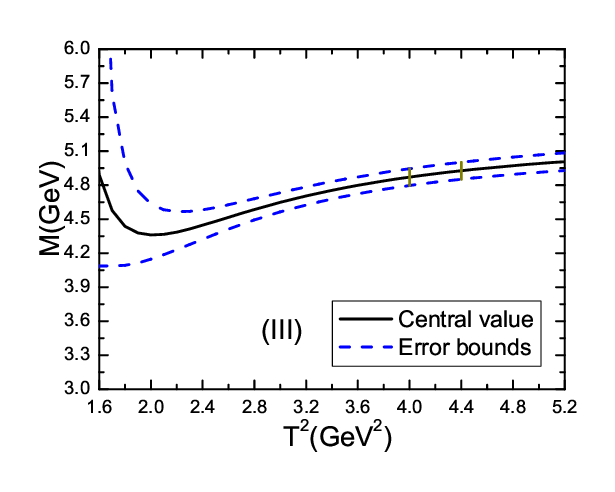}
\includegraphics[totalheight=5cm,width=7cm]{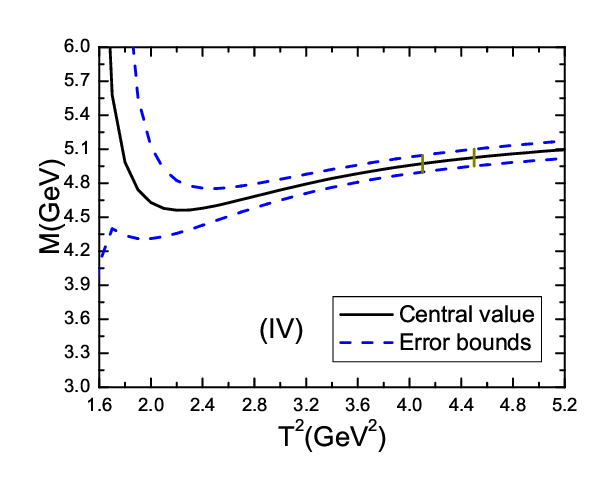}
\includegraphics[totalheight=5cm,width=7cm]{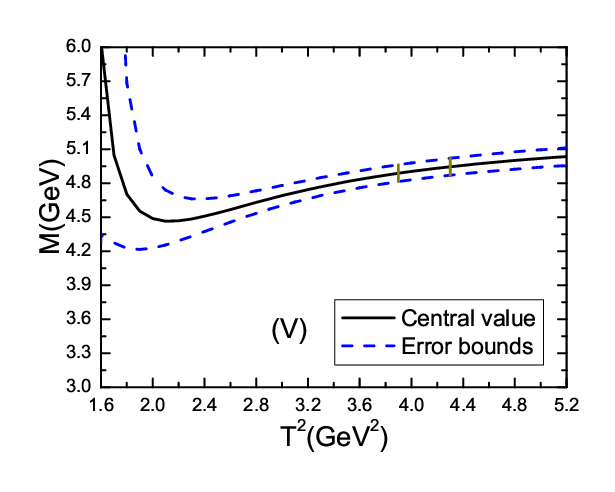}
\includegraphics[totalheight=5cm,width=7cm]{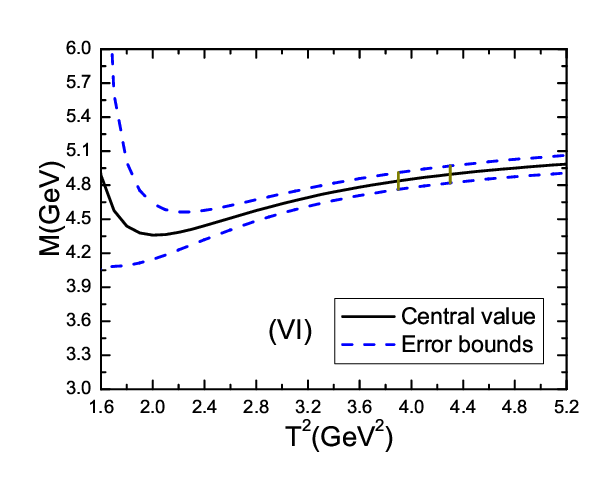}
\includegraphics[totalheight=5cm,width=7cm]{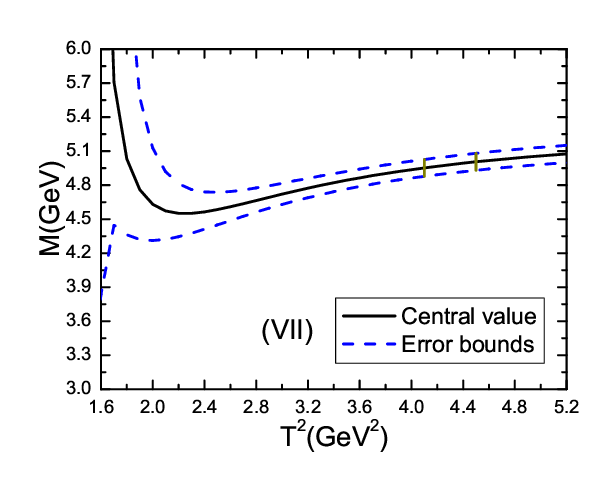}
  \caption{ The masses  with variations of the  Borel parameters $T^2$ for  the $sssc\bar{c}$  pentaquark states with negative parity, where the (I), (II), (III), (IV), (V),  (VI) and (VII)  denote the    $[ss][sc]\bar{c}$ ($1$, $1$, $0$, $\frac{1}{2}$),
$[ss][sc]\bar{c}$ ($1$, $0$, $1$, $\frac{1}{2}$),
$[ss][sc]\bar{c}$ ($1$, $0$, $1$, $\frac{3}{2}$),
$[ss][sc]\bar{c}$ ($1$, $1$, $2$, $\frac{3}{2}$)${}_2$,
$[ss][sc]\bar{c}$ ($1$, $1$, $2$, $\frac{3}{2}$)${}_3$,
 $[ss][sc]\bar{c}$ ($1$, $0$, $1$, $\frac{5}{2}$) and $[ss][sc]\bar{c}$ ($1$, $1$, $2$, $\frac{5}{2}$) pentaquark states, respectively. }\label{mass-1-fig}
\end{figure}

We consider   all uncertainties  of the input   parameters,
and obtain  the masses and pole residues of
 the diquark-diquark-antiquark type  $sssc\bar{c}$  pentaquark states with negative parity, and show them plainly in Fig.\ref{mass-1-fig} and Table \ref{mass-Pcs}. From Tables \ref{Borel}-\ref{mass-Pcs}, we observe   that the predicted pentaquark masses and the pertinent  energy scales of the QCD spectral densities satisfy the modified energy scale formula
 $\mu =\sqrt{M^2_{P}-(2{\mathbb{M}}_c)^2}-3{\mathbb{M}}_s$.
 Without adopting  the (modified) energy scale formula, we could only acquire    bad convergent behavior of the OPE and small pole contributions in the QCD sum rules for the multiquark states \cite{WangZG-IJMPA-3-scheme}.

In Fig.\ref{mass-1-fig}, we plot the masses of the $sssc\bar{c}$ pentaquark states with negative parity via variations of the Borel parameters at rather large intervals, the regions between the two short perpendicular   lines are the Borel windows, where  flat platforms appear obviously.

\begin{figure}
\centering
\includegraphics[totalheight=6cm,width=7cm]{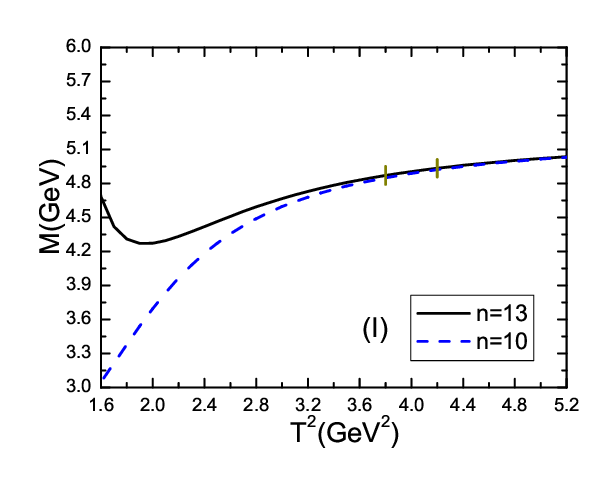}
\includegraphics[totalheight=6cm,width=7cm]{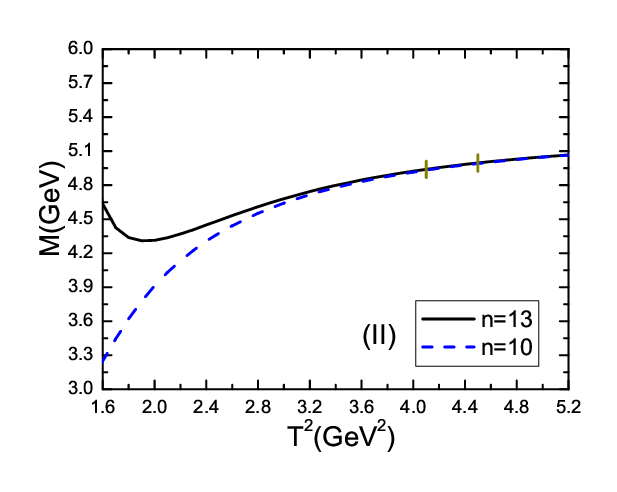}
  \caption{ The masses  with variations of the  Borel parameters $T^2$ for  the $sssc\bar{c}$  pentaquark states with truncation of the OPE up to the vacuum condensates of dimensions 10 and 13, where the (I) and (II)  denote the    $[ss][sc]\bar{c}$ ($1$, $1$, $0$, $\frac{1}{2}$) and
$[ss][sc]\bar{c}$ ($1$, $0$, $1$, $\frac{1}{2}$) pentaquark states, respectively. }\label{mass-D10-fig}
\end{figure}

\begin{figure}
\centering
\includegraphics[totalheight=6cm,width=7cm]{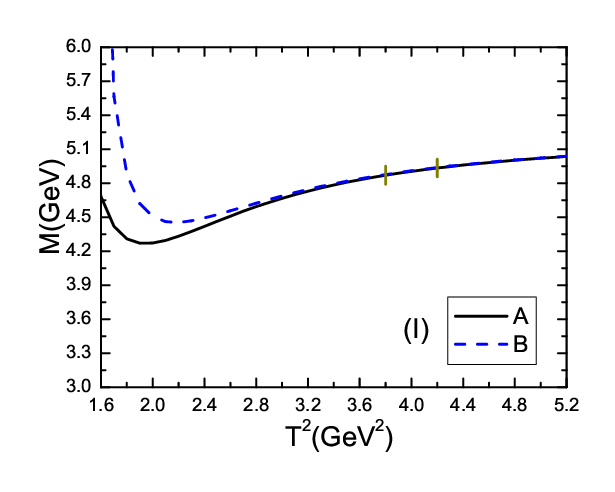}
\includegraphics[totalheight=6cm,width=7cm]{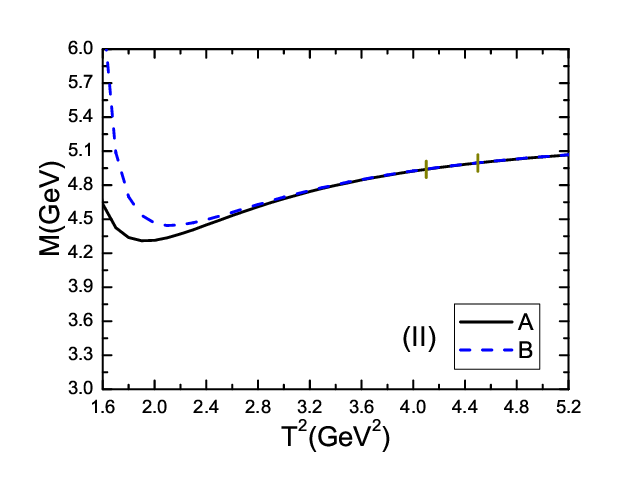}
  \caption{ The masses  with variations of the  Borel parameters $T^2$ for  the $[ss][sc]\bar{c}$ ($1$, $1$, $0$, $\frac{1}{2}$) (I) and
$[ss][sc]\bar{c}$ ($1$, $0$, $1$, $\frac{1}{2}$) (II)  pentaquark states, where the $A$ and $B$ denote the thresholds $4m_c^2$ and $(2m_c+3m_s)^2$, respectively. }\label{mass-3ms-fig}
\end{figure}

In Table \ref{mass-Pcs}, we also present  our previous predictions in Refs.\cite{WangZG-EPJC-1509-12,WangZG-NPB-1512-32},  the new analysis enlarges the pentaquark  masses about $200-300\,\rm{MeV}$. The present calculations are more robust, as we take account of more vacuum condensates in the OPE to obtain more flat platforms, which weakens the dependence on the Borel parameters significantly.

In Fig.\ref{mass-D10-fig}, as an example, we take the parameters in Table 2, and plot the predicted masses  with variations of the  Borel parameters $T^2$ for  the $[ss][sc]\bar{c}$ ($1$, $1$, $0$, $\frac{1}{2}$) and
$[ss][sc]\bar{c}$ ($1$, $0$, $1$, $\frac{1}{2}$)  pentaquark states with truncation of the OPE up to the vacuum condensates of dimensions 10 and 13, respectively.
In the case of the truncation of the OPE with $n=10$,  the predicted masses increase monotonously and quickly with increase of the Borel parameters before reaching the Borel windows, which are indicted by two short perpendicular lines. In the Borel windows, the two truncations make almost no difference to the pentaquark masses.
However, direct calculations indicate that such Borel windows do not survive in the case of the truncation of the OPE with $n=10$, we obtain quite different Borel windows and quite different pentaquark masses \cite{WangZG-EPJC-1509-12,WangZG-NPB-1512-32}. The vacuum condensates with the dimensions $11$ and $13$ are associated with  inverse powers of the Borel parameters, i.e. $\frac{1}{T^2}$, $\frac{1}{T^4}$, $\frac{1}{T^6}$ or $\frac{1}{T^8}$ (see Appendix), and manifest themselves at small values of  $T^2$. Thus they play an important role in determining the Borel windows, while in the Borel windows, they play a tiny role. If we retain the Borel windows but neglect the contributions of the vacuum condensates with dimensions larger than $10$, all the predicted pentaquark masses undergo a systematic error $\delta M_P=-0.01\,\rm{GeV}$.

We usually carry out the OPE with the help of the full light quark propagator \cite{PRT85,WangHuang3900,Pascual-1984},
\begin{eqnarray}\label{q-full-propagator}
S_{ij}(x)&=& \frac{i\delta_{ij}\!\not\!{x}}{ 2\pi^2x^4}
-\frac{\delta_{ij}m_q}{4\pi^2x^2}-\frac{\delta_{ij}\langle
\bar{q}q\rangle}{12} +\frac{i\delta_{ij}\!\not\!{x}m_q
\langle\bar{q}q\rangle}{48}-\frac{\delta_{ij}x^2\langle \bar{q}g_s\sigma Gq\rangle}{192}+\frac{i\delta_{ij}x^2\!\not\!{x} m_q\langle \bar{q}g_s\sigma
 Gq\rangle }{1152}\nonumber\\
&& -\frac{ig_s G^{a}_{\alpha\beta}t^a_{ij}(\!\not\!{x}
\sigma^{\alpha\beta}+\sigma^{\alpha\beta} \!\not\!{x})}{32\pi^2x^2} -\frac{\delta_{ij}x^4\langle \bar{q}q \rangle\langle g_s^2 GG\rangle}{27648}-\frac{1}{8}\langle\bar{q}_j\sigma^{\mu\nu}q_i \rangle \sigma_{\mu\nu}  +\cdots \, ,
\end{eqnarray}
 where $q=u$, $d$ or $s$. In Eq.\eqref{q-full-propagator}, we take the light quark mass $m_q$ as a small quantity, and take account of its contribution perturbatively. Thus, the dispersion relation in Eq.\eqref{QCD-rho} corresponds to the limit $m_q \to 0$ \cite{WZG-penta-IJMPA,WangZG-Pc12-JpsiLambda,WangZG-Pc12-Jpsip,
WangZG-Pc12-JpsiXi,WangZG-Pc12-JpsiSgm,WangZG-Pc12-JpsiXi-10}. If we want to take account of the light quark mass $m_q$ as a finite quantity, we should write down the full propagator in Eq.\eqref{q-full-propagator} in the momentum space,
\begin{eqnarray}
S_{ij}(k)&=&\delta_{ij}\frac{\!\not\!{k}+m_q}{k^2-m_q^2}+\cdots\, ,
\end{eqnarray}
the finite $m_q^2$ in the denominator makes  calculations extremely complicated. We can take account of the collective  effects of the finite mass $m_q$ roughly by taking  a simple replacement,
\begin{eqnarray}
\int_{4m_c^2}^{s_0}ds &\to & \int_{(2m_c+3m_s)^2}^{s_0}ds \, ,
\end{eqnarray}
 in Eq.\eqref{QCDSR} and Eq.\eqref{QCDSR-mass}.
Direct calculations indicate that the Borel windows survive and such a replacement only leads to a systematic error $\delta M_P=+(0.00\sim0.01)\,\rm{GeV}$.

In Fig.\ref{mass-3ms-fig}, as an example, we plot the masses  with variations of the  Borel parameters $T^2$ for  the $[ss][sc]\bar{c}$ ($1$, $1$, $0$, $\frac{1}{2}$)  and
$[ss][sc]\bar{c}$ ($1$, $0$, $1$, $\frac{1}{2}$)  pentaquark states with the thresholds $4m_c^2$ and $(2m_c+3m_s)^2$, respectively. From the figure, we can see explicitly that such a replacement manifests itself only at the region $T^2\leq 2.4\,\rm{GeV}^2$, in the Borel windows, which are indicted by two short perpendicular lines, the replacement almost makes no difference to the pentaquark masses.

Now we compare the present calculations to the existing works. In Ref.\cite{Zong-penta-sss-PLB-19}, Meng et al studied  the compact $sssc\bar c$ pentaquark resonances in a non-relativistic potential quark model considering five sets of Jacobi coordinate systems. They  predicted that the lowest pentaquark states with the spin-parity $J^P={\frac{1}{2}}^-$, ${\frac{3}{2}}^-$ and ${\frac{5}{2}}^-$ have  masses $4855\,\rm{MeV}$, $4753\,\rm{MeV}$ and $4873\,\rm{MeV}$, respectively, which are compatible with the present calculations except for the $4753\,\rm{MeV}$.

In Ref.\cite{Li-penta-sss-PRD-23}, Li et al  studied the mass spectrum of the S-wave hidden-charm pentaquark states in a color-magnetic interaction model with the color singlet-singlet plus octet-octet wave-functions, and obtained the ground state masses $4.6\sim 4.8\,\rm{GeV}$ for the $sssc\bar{c}$ states, which are smaller than the present calculations significantly.

On the other hand, in the molecule scenario, the predicted masses usually lie at the nearby two-particle thresholds. For example, in the one-boson-exchange model,  Wang et al observed   that the S-wave $\Omega_{c}\bar D_s^*$ ($\Omega_{c}^{*}\bar D_s^*$) state with the spin-parity $J^P={\frac{3}{2}}^{-}$ (${\frac{5}{2}}^{-}$)  is a possible  candidate of the hidden-charm molecule  with triple-strange \cite{WangFL-mole-sss-PRD-21}.

In Ref.\cite{Clymton-mole-sss-PRD-25}, Clymton, Kim and Mart investigated possible existence of the triple-strange hidden-charm pentaquark states in the off-shell coupled-channel formalism, and  observed two triple-strange hidden-charm pentaquark states $P_{csss}(4787)$ and $P_{csss}(4841)$  with the spin-parity $J^P={\frac{1}{2}}^-$, which lie just below the $\bar{D}_s^*\Omega_c^*$ threshold $4878\,\rm{MeV}$.

However, in Ref.\cite{Oset-mole-sss-PRD-24},  Roca, Song and Oset  explored  possible dynamically generated double- and triple-strange pentaquark-type states using the unitarized  coupled-channel processes, and observed that the interaction is not strong enough to generate pole in the triple-strange channels.

All the theoretical predictions are model-dependent in one way or the other, and differ from each other in one way or the other. Furthermore,  there exists no experimental data on the hidden-charm pentaquark states with triple-strange, thus  no definite conclusion could be obtained at the present time.

Unostentatiously, we look forward to observe the  $P_{css}$ and $P_{csss}$ states with strangeness $S=-2$ and $-3$ respectively in the weak decays of the ground state bottom baryons,
\begin{eqnarray}
\Omega_b^{-}(bss)&\to& P_{css}^-\,(\mathbf{10}) \bar{K}^0 \to J/\psi \Xi^{*-}\, \bar{K}^0\, ,\nonumber\\
&\to& P_{css}^-\,(\mathbf{8}) \bar{K}^0 \to J/\psi \Xi^{-}\, \bar{K}^0\, ,\nonumber\\
&\to&P_{csss}^-(\mathbf{10})\,\phi \to J/\psi \Omega^{-}  \phi \, ,
\end{eqnarray}
via the CKM favored decay $b \to c\bar{c}s$ at the level of  quark-gluon degrees of freedom. The predictions for the $P_{css}$ states in the light flavor $\mathbf{8}$ and $\mathbf{10}$ representations are given clearly in Refs.\cite{WangZG-Pc12-JpsiXi} and \cite{WangZG-Pc12-JpsiXi-10}, respectively.
The $P_c(4312)$, $P_c(4380)$,  $P_c(4440)$, $P_c(4457)$  and $P_{cs}(4338)$ were observed in the $J/\psi p$ and $J/\psi \Lambda$ invariant mass distributions, respectively, and evidences of the   $P_c(4337)$ and $P_{cs}(4459)$   were observed in the $J/\psi p$  and $J/\psi \Lambda$ invariant mass distributions, respectively. Confirmations of the $P_c(4337)$ and $P_{cs}(4459)$
and observations of other hidden-charm pentaquark candidates in the  $J/\psi \Delta$, $J/\psi\Xi$, $J/\psi\Sigma$, $J/\psi\Xi^*$, $J/\psi\Sigma^*$ and $J/\psi\Omega$ invariant mass distributions, especially in the $J/\psi\Omega$ invariant mass distributions which have strangeness $S=-3$, would shed light on the nature of exotic states.

We would like to take the pole residues as  input parameters to compute the following   two-body strong decays,
 \begin{eqnarray}\label{decay-Pcsss}
P_{csss}&\to&  \bar{D}_s\Omega_c(4664)\, , \, \bar{D}_s\Omega_c^*(4735)\, , \,\bar{D}_s^*\Omega_c(4807)\, ,  \,\bar{D}_s^*\Omega_c^*(4878)\, ,  \,J/\psi \Omega(4769) \, , \, \eta_c \Omega(4656) \, ,\nonumber\\
\end{eqnarray}
 with the traditional  QCD sum rules to obtain the partial decay widths as guide and pick up the optimal decay channels to hunt for the $P_{csss}$ states experimentally, where we write down  thresholds of the meson-baryon pairs in the bracket and with the unit of $\rm{MeV}$.
The central values of the $P_{csss}$ states lie above the corresponding  meson-baryon thresholds except for the $[ss][sc]\bar{c}$ ($1$, $0$, $1$, $\frac{5}{2}$) state, which lies slightly below the $\bar{D}_s^*\Omega_c^*$ threshold.

In fact, the decays shown in Eq.\eqref{decay-Pcsss} can take place if they are kinematically  allowed, irrespective of assigning  the $P_{csss}$ states in the pentaquark or molecule  scenarios. In the present work, we would like to adopt the predictions  of the QCD sum rules as we focus on the QCD sum rules, where the $sssc\bar{c}$ molecular states lie below the corresponding two-particle thresholds, the decays to their constituents $\bar{D}_s\Omega_c$, $\bar{D}_s\Omega_c^*$, $\bar{D}_s^*\Omega_c$ or $\bar{D}_s^*\Omega_c^*$  are kinematically  forbidden \cite{WangXW-penta-mole-sss}.

In Ref.\cite{Azizi-penta-sss-PRD-23},   Azizi, Sarac and Sundu  studied the two-body strong decays $P_{csss} \to  J/\psi\Omega $ with the three-point QCD sum rules using our old calculations \cite{WangZG-EPJC-1509-12,WangZG-NPB-1512-32}, and obtained the partial decay widths $201.4\pm 82.5~\mathrm{MeV}$  and $252.5\pm 116.7~\mathrm{MeV}$  for the $[ss][sc]\bar{c}$ ($1$, $1$, $0$, $\frac{1}{2}$)  and
$[ss][sc]\bar{c}$ ($1$, $0$, $1$, $\frac{1}{2}$)  pentaquark states, respectively, which are much larger than the ones in the molecule scenario \cite{WangFL-mole-sss-PRD-21,Clymton-mole-sss-PRD-25}.
In calculations, Azizi, Sarac and Sundu did not subtract  contaminations from the higher  resonances and continuum states in the pentaquark channels, which is expect to lead to larger hadronic coupling constants therefore larger partial decay widths.

All in all, more experimental data and theoretical works are still needed,  observations of the $P_{csss}$ states would shed light on the nature of  exotic states and provide an excellent opportunity to distinguish the diquark-diquark-antiquark type pentaquark states and meson-baryon molecular states.

\section{Conclusion}
 In this work, we construct the diquark-diquark-antiquark type currents to interpolate  the hidden-charm pentaquark states with strangeness $S=-3$  in details. We compute the  vacuum condensates up to dimension $13$ in the OPE consistently, get the QCD spectral representation and distinguish  contributions of the
 $sssc\bar{c}$ pentaquark  states  with negative and positive parities respectively,
adopt  the modified energy scale formula $\mu=\sqrt{M_{P}^2-(2{\mathbb{M}}_c)^2}-3{\mathbb{M}}_s$ to choose  the pertinent   energy scales. Finally, we obtain the mass spectrum of the $sssc\bar{c}$ pentaquark states with the quantum numbers $IJ^{P}=0{\frac{1}{2}}^-$, $0{\frac{3}{2}}^-$, $0{\frac{5}{2}}^-$, which serves as a guide to hunt for the hidden-charm pentaquark states with strangeness $S=-3$ in the exclusive decays $\Omega_b^- \to P_{csss}^-\phi \to J/\psi \Omega^-  \phi $. Observations of the $P_{csss}$ states would shed light on the nature of  exotic states and provide an excellent opportunity to distinguish the diquark-diquark-antiquark type pentaquark states from meson-baryon molecular states

\section*{Appendix}
The explicit expressions of the QCD spectral densities for the current $J^1(x)$,
\begin{eqnarray}
\rho^{1/0}_{QCD}(s)&=&\sum_k \rho^{1/0}_{k}(s)\, ,
\end{eqnarray}
\begin{eqnarray}
\rho^{0}_{k}(s)&=&m_c\,\tilde{\rho}^{0}_{k}(s)\, ,
\end{eqnarray}
with $k=0$, $3$, $4$, $5$, $6$, $7$, $8$, $9$, $10$, $11$, $13$,
where
\begin{eqnarray}
\rho^{1}_{0}(s)&=& \frac{1}{61440\pi^8}\int dydz\, zy(1-y-z)^{4} (s-\bar{m}_{c}^{2})^{4}(8s-3\bar{m}_{c}^{2})\exp\left(-\frac{s}{T^{2}}\right)\nonumber\\
&&+\frac{m_s m_c}{6144\pi^8}\int dy  dz\,z(1-y-z)^3	(s-\bar{m}_c^2)^4
\exp\left(-\frac{s}{T^2}\right) \, ,
\end{eqnarray}

\begin{eqnarray}
\rho^{1}_{3}(s)&=&-\frac{ m_{c}\langle\bar{s}s\rangle}{384\pi^{6}}
\int dydz	\, z(1-y-z)^{2} (s-\bar{m}_{c}^{2})^{3}		
\exp\left(-\frac{s}{T^{2}}\right) \nonumber\\	 &&+\frac{m_s\langle\bar{s}s\rangle}{128\pi^6} \int dydz\,zy(1-y-z)^2
(s-\bar{m}_c^2)^2 (2s-\bar{m}_c^2)\exp\left(-\frac{s}{T^2}\right) \, ,
\end{eqnarray}

\begin{eqnarray}
\rho^{1}_{4}(s)&=&	-\frac{m_{c}^{2}\left\langle\tilde{\alpha} GG\right\rangle}{9216\pi^{6}}
 \int dydz\frac{z(1-y-z)^{4}}{y^{2}}\left(s-\bar{m}_{c}^{2}\right)	 \left(5s-3\bar{m}_{c}^{2}\right)\exp\left(-\frac{s}{T^{2}}\right) \nonumber  \\
&&-\frac{\left\langle\tilde{\alpha} GG\right\rangle}{2048\pi^{6}}
\int dydz\, zy(1-y-z)^{2}\left(s-\bar{m}_{c}^{2}\right)^{2} \left(2s-\bar{m}_{c}^{2}\right) \exp\left(-\frac{s}{T^{2}}\right) \nonumber  \\
&&+\frac{\left\langle\tilde{\alpha} GG\right\rangle}{3072\pi^{6}}
\int dydz\, z(1-y-z)^{3}\left(s-\bar{m}_{c}^{2}\right)^{2} \left(2s-\bar{m}_{c}^{2}\right)	\exp\left(-\frac{s}{T^{2}}\right) \nonumber  \\
&&-\frac{m_s m_c^3 \left\langle \tilde{\alpha}GG \right\rangle}{4608\pi^6}
\int dy dz\,\frac{(1-y-z)^3}{y^2}\left( 1+\dfrac{z}{y}\right)			 (s-\bar{m}_c^2)\exp\left(-\frac{s}{T^2}\right) \nonumber \\
&&+\frac{m_s m_c \left\langle \tilde{\alpha}GG \right\rangle}{3072\pi^6}
\int dydz\,\frac{z(1-y-z)^3}{y^2}(s-\bar{m}_c^2)^2	 \exp\left(-\frac{s}{T^2}\right)
\nonumber  \\
&& -\frac{7m_s m_c \left\langle \tilde{\alpha}GG \right\rangle}{4096\pi^6}
\int dydz\,	z(1-y-z)(s-\bar{m}_c^2)^2\exp\left(-\frac{s}{T^2}\right) \nonumber \\
&&+ \frac{m_s m_c \left\langle \tilde{\alpha}GG \right\rangle}{1024\pi^6}
\int dydz\, \frac{z(1-y-z)^2}{y}(s-\bar{m}_c^2)^2		 \exp\left(-\frac{s}{T^2}\right)\, ,		
\end{eqnarray}

\begin{eqnarray}
\rho^{1}_{5}(s)&=&\frac{3m_{c}\langle\bar{s}Gs\rangle}{512\pi^{6}} \int dy dz	\, z(1-y-z) (s-\bar{m}_{c}^{2})^{2}	 \exp\left(-\frac{s}{T^{2}}\right) \nonumber\\ &&+\frac{m_s \langle\bar{s}Gs\rangle}{512\pi^6}\int dydz\,zy(1-y-z)
(s-\bar{m}_c^2) (5s-3\bar{m}_c^2)\exp\left(-\frac{s}{T^2}\right) \nonumber\\
&& - \frac{m_s \langle\bar{s}Gs\rangle}{512\pi^6}\int dydz\,z(1-y-z)^2
(s-\bar{m}_c^2) (5s-3\bar{m}_c^2)\exp\left(-\frac{s}{T^2}\right)  \, ,	 \end{eqnarray}

\begin{eqnarray}
\rho^{1}_{6}(s)&=&\frac{\langle\bar{s}s\rangle^{2}}{48\pi^{4}}
 \int dy dz\,  zy(1-y-z) (s-\bar{m}_{c}^{2})(5s-3\bar{m}_{c}^{2})		 \exp\left(-\frac{s}{T^{2}}\right) \nonumber\\
 &&+\frac{5m_s m_c \langle\bar{s}s\rangle^2}{24\pi^4}\int dydz\,
z(s-\bar{m}_c^2)\exp\left(-\frac{s}{T^2}\right)\, ,
\end{eqnarray}

\begin{eqnarray}
\rho^{1}_{7}(s)&=&\frac{m_{c}^{3}\langle\bar{s}s\rangle
\left\langle\tilde{\alpha} GG\right\rangle}{1152\pi^{4}}
 \int dy dz	\left(1+\frac{z}{y}\right)\frac{(1-y-z)^{2}}{y^2}
\exp\left(-\frac{s}{T^{2}}\right) \nonumber  \\
&&-\frac{m_{c}\langle\bar{s}s\rangle\left\langle\tilde{\alpha} GG\right\rangle}{384\pi^{4}}
\int dydz \frac{z(1-y-z)^{2}}{y^{2}}\left(s-\bar{m}_{c}^{2}\right)
\exp\left(-\frac{s}{T^{2}}\right)  \nonumber  \\			
&&+\frac{17m_{c}\langle\bar{s}s\rangle\left\langle\tilde{\alpha} GG\right\rangle}{4608\pi^{4}}
\int dydz\, z\left(s-\bar{m}_{c}^{2}\right)
\exp\left(-\frac{s}{T^{2}}\right) \nonumber  \\
&&-\frac{m_{c}\langle\bar{s}s\rangle
\left\langle\tilde{\alpha} GG\right\rangle}{192\pi^{4}}
\int dydz\frac{z(1-y-z)}{y}	\left(s-\bar{m}_{c}^{2}\right)
\exp\left(-\frac{s}{T^{2}}\right) \nonumber  \\
&&-\frac{m_s m_c^2\langle\bar{s}s\rangle \left\langle\tilde{\alpha}GG\right\rangle}{192 \pi^4}
\int dydz\,\frac{z(1-y-z)^2}{y^2}\left[ 1+ \frac{s}{3}\delta(s-\bar{m}_c^2)\right]
\exp\left(-\frac{s}{T^2}\right) \nonumber\\
&& - \frac{7m_s \langle\bar{s}s\rangle \left\langle \tilde{\alpha}GG \right\rangle}{2304\pi^4}\int dydz\,zy(4s-3\bar{m}_c^2)
\exp\left(-\frac{s}{T^2}\right) \nonumber\\
&&+ \frac{5m_s \langle\bar{s}s\rangle \left\langle \tilde{\alpha}GG \right\rangle}{768\pi^4}\int dydz\,	z(1-y-z)(4s-3\bar{m}_c^2)
\exp\left(-\frac{s}{T^2}\right) \, ,
\end{eqnarray}

\begin{eqnarray}
\rho^{1}_{8}(s)&=&-\frac{5\langle\bar{s}s\rangle\langle\bar{s}Gs\rangle}{192\pi^{4}}
\int dydz\, zy (4s-3\bar{m}_{c}^{2})	 \exp\left(-\frac{s}{T^{2}}\right)\nonumber\\
&&+\frac{\langle\bar{s}s\rangle\langle\bar{s}Gs\rangle}{96\pi^{4}}
\int dydz\, z(1-y-z)\left(4s-3\bar{m}_{c}^{2}\right)
\exp\left(-\frac{s}{T^{2}}\right) \nonumber  \\
&&-\frac{121m_s m_c \langle\bar{s}s\rangle\langle\bar{s}Gs\rangle}{1152\pi^4}
\int dy\,(1-y)\exp\left(-\frac{s}{T^2}\right)\nonumber \\
&&+ \frac{m_s m_c \langle\bar{s}s\rangle\langle\bar{s}Gs\rangle}{24\pi^4}
\int dydz\, \frac{z}{y} \exp\left(-\frac{s}{T^2}\right) \, ,
\end{eqnarray}

\begin{eqnarray}
\rho^{1}_{9}(s)&=&-\frac{m_{c}\langle\bar{s}s\rangle^{3}}{9\pi^{2}}
 \int dy (1-y) 	\exp\left(-\frac{s}{T^{2}}\right)\nonumber\\	
&&+ \frac{m_s \langle\bar{s}s\rangle^3}{12 \pi^2}\int dy\,y(1-y)
\left[1+\frac{s}{3}\delta(s-\tilde{m}_c^2)\right]
\exp\left(-\frac{s}{T^2}\right)\, ,
\end{eqnarray}

\begin{eqnarray}
\rho^{1}_{10}(s)&=&\frac{3\langle\bar{s}Gs\rangle^{2}}{256\pi^{4}}
\int dy \, y(1-y) 	\left[1+\frac{s}{3}\delta(s-\tilde{m}_{c}^{2})
\right]	\exp\left(-\frac{s}{T^{2}}\right) \nonumber\\		
&&-\frac{m_{c}^{2}\langle\bar{s}s\rangle^{2}\left\langle\tilde{\alpha} GG\right\rangle}{108\pi^{2}}
 \int dy dz	\frac{z(1-y-z)}{y^{2}} \left(1+\frac{s}{2T^{2}}\right)
\delta\left(s-\bar{m}_{c}^{2}\right)\exp\left(-\frac{s}{T^{2}}\right) \nonumber  \\
&&-\frac{\langle\bar{s}s\rangle^{2}\left\langle\tilde{\alpha} GG\right\rangle}{384\pi^{2}}
\int dydz\, z \left[1+\frac{s}{3}\delta\left(s-\bar{m}_{c}^{2}\right)\right]
\exp\left(-\frac{s}{T^{2}}\right) \nonumber  \\
&&-\frac{5\langle\bar{s}Gs\rangle^{2}}{512\pi^{4}}
 \int dydz\, z \left[1+\frac{s}{3}\delta\left(s-\bar{m}_{c}^{2}\right)\right]
\exp\left(-\frac{s}{T^{2}}\right) \nonumber  \\
&&+\frac{\langle\bar{s}s\rangle^{2}\left\langle\tilde{\alpha} GG\right\rangle}{144\pi^{2}}
\int dy\,y(1-y) \left[1+\frac{s}{3}\delta\left(s-\tilde{m}_{c}^{2}\right)\right]
\exp\left(-\frac{s}{T^{2}}\right) \nonumber  \\	
&& + \frac{31m_s m_c \langle\bar{s}Gs\rangle^2}{1152\pi^4}\int dy\,
(1-y)\left(1+\frac{s}{2 T^2} \right)		 \delta(s-\tilde{m}_c^2)\exp\left(-\frac{s}{T^2}\right) \nonumber\\
&& -\frac{5m_s m_c^3 \langle\bar{s}s\rangle^2 \left\langle \tilde{\alpha}GG \right\rangle}{432\pi^2 T^2}\int dydz\,\frac{1}{y^2}\left(1+\frac{z}{y}\right)
\delta(s-\bar{m}_c^2)\exp\left(-\frac{s}{T^2}\right)\nonumber \\
&&+ \frac{5m_s m_c \langle\bar{s}s\rangle^2 \left\langle \tilde{\alpha}GG \right\rangle}{144\pi^2}\int dydz\, \frac{z}{y^2}
\delta(s-\bar{m}_c^2)\exp\left(-\frac{s}{T^2}\right) \nonumber\\
&&- \frac{m_s m_c \langle\bar{s}s\rangle^2 \left\langle \tilde{\alpha}GG \right\rangle}{288\pi^2}\int dy\, \frac{(1-y)}{y}
\delta(s-\tilde{m}_c^2)\exp\left(-\frac{s}{T^2}\right)\nonumber \\
&&-\frac{m_s m_c \langle\bar{s}Gs\rangle^2}{96\pi^4}\int dy\,
\frac{(1-y)}{y}	\delta(s-\tilde{m}_c^2)\exp\left(-\frac{s}{T^2}\right) \nonumber\\
&&+\frac{11 m_s m_c \langle\bar{s}s\rangle^2 \left\langle \tilde{\alpha}GG\right\rangle}{432 \pi^2}\int dy\,(1-y)
\left(1+\frac{s}{2 T^2} \right)		 \delta(s-\tilde{m}_c^2)\exp\left(-\frac{s}{T^2}\right)\, ,
\end{eqnarray}

\begin{eqnarray}
\rho^{1}_{11}(s)&=&\frac{m_{c}\langle\bar{s}s\rangle^{2}\langle\bar{s}Gs\rangle}{6\pi^{2}}
\int dy\, (1-y) \left(1+\frac{s}{2T^{2}}\right)\delta(s-\tilde{m}_{c}^{2})
\exp\left(-\frac{s}{T^{2}}\right) \nonumber \\	
&&-\frac{m_{c}\langle\bar{s}s\rangle^{2}\langle\bar{s}Gs\rangle}{36\pi^{2}}
\int dy \frac{(1-y)}{y}	\delta\left(s-\tilde{m}_{c}^{2}\right) \exp\left(-\frac{s}{T^{2}}\right) \nonumber  \\	
&&-\frac{ m_s \langle\bar{s}s\rangle^2\langle\bar{s}Gs\rangle}{9\pi^2}
\int dy\,y(1-y) \left(1+\frac{2s}{3T^2}+\frac{s^2}{6T^4}\right)
\delta(s-\tilde{m}_c^2)\exp\left(-\frac{s}{T^2}\right) \nonumber\\
&& + \frac{m_s \langle\bar{s}s\rangle^2\langle\bar{s}Gs\rangle}{72\pi^2}
\int dy\,(1-y)\left(1+\frac{s}{2T^2} \right)
\delta(s-\tilde{m}_c^2)\exp\left(-\frac{s}{T^2}\right) \, ,
\end{eqnarray}

\begin{eqnarray}
\rho^{1}_{13}(s)&=&-\frac{m_{c}\langle\bar{s}s\rangle\langle\bar{s}Gs\rangle^{2}}{24\pi^2T^2}
\int dy \, (1-y)\left(	1+\frac{s}{T^{2}}+\frac{s^2}{2T^{4}}\right)
\delta\left(s-\tilde{m}_{c}^{2}\right)\exp\left(-\frac{s}{T^{2}}\right) \nonumber \\
&&+\frac{m_{c}\langle\bar{s}s\rangle\langle\bar{s}Gs\rangle^{2}}{72\pi^{2}T^{2}}
\int dy\, \frac{(1-y)}{y} \left(1+\frac{s}{T^{2}}\right)
\delta\left(s-\tilde{m}_{c}^{2}\right)\exp\left(-\frac{s}{T^{2}}\right) \nonumber  \\
&&+\frac{m_{c}^{3}\langle\bar{s}s\rangle^{3}
\left\langle\tilde{\alpha} GG\right\rangle}{162T^{4}} \int dy \frac{1}{y^3}		 \delta\left(s-\tilde{m}_{c}^{2}\right)\exp\left(-\frac{s}{T^{2}}\right) \nonumber  \\
&&-\frac{m_{c}\langle\bar{s}s\rangle^{3}
\left\langle\tilde{\alpha} GG\right\rangle}{54T^{2}}
\int dy\,\frac{(1-y)}{y^{2}}\delta\left(s-\tilde{m}_{c}^{2}\right) \exp\left(-\frac{s}{T^{2}}\right) \nonumber\\
&&-\frac{m_{c}\langle\bar{s}s\rangle^{3}
\left\langle\tilde{\alpha} GG\right\rangle}{54T^{2}}
\int dy \, (1-y) \left(1+\frac{s}{T^{2}}+\frac{s^2}{2T^{4}}\right)
\delta\left(s-\tilde{m}_{c}^{2}\right)\exp\left(-\frac{s}{T^{2}}\right) \nonumber \\
&&+\frac{7m_s \langle\bar{s}s\rangle\langle\bar{s}Gs\rangle^2}{288\pi^2 T^2}
\int dy\,y(1-y) \left(1+\frac{s}{T^2}+\frac{s^2}{2T^4}+\frac{s^3}{6T^6}\right)
\delta(s-\tilde{m}_c^2)\exp\left(-\frac{s}{T^2}\right) \nonumber\\
&& -\frac{m_s m_c^2 \langle\bar{s}s\rangle^3 \left\langle \tilde{\alpha}GG \right\rangle}{324 T^6}	\int dy\, \frac{(1-y)}{y^2} \,s\,
 \delta(s-\tilde{m}_c^2)\exp\left(-\frac{s}{T^2}\right) \nonumber\\
&& - \frac{5m_s \langle\bar{s}s\rangle\langle\bar{s}Gs\rangle^2}{864\pi^2 T^2}
\int dy\,(1-y)\left( 1+\frac{s}{T^2}+\frac{s^2}{2 T^4} \right)
\delta(s-\tilde{m}_c^2)\exp\left(-\frac{s}{T^2}\right) \nonumber\\
&& + \frac{m_s \langle\bar{s}s\rangle^3 \left\langle \tilde{\alpha}GG\right\rangle}{108 T^2}	\int dy\,y(1-y)
\left( 1+\frac{s}{T^2}+\frac{s^2}{2 T^4}+\frac{s^3}{6 T^6} \right)
\delta(s-\tilde{m}_c^2)\exp\left(-\frac{s}{T^2}\right)\, ,
\end{eqnarray}

\begin{eqnarray}
\tilde{\rho}_0^0(s)&=&\frac{1}{61440\pi^{8}} \int dydz
\, y(1-y-z)^{4} \left(s-\bar{m}_{c}^{2}\right)^{4}\left(7s-2\bar{m}_{c}^{2}\right)
\exp\left(-\frac{s}{T^{2}}\right) \nonumber \\
&&+\frac{m_s m_c}{6144\pi^8} \int dydz\,(1-y-z)^3
\left(s-\bar{m}_c^2\right)^4\exp\left(-\frac{s}{T^2}\right) \, ,
\end{eqnarray}

\begin{eqnarray}
\tilde{\rho}_3^0(s)&=&-\frac{m_{c}\langle\bar{s}s\rangle}{384\pi^{6}}
\int dydz\, (1-y-z)^{2}\left(s-\bar{m}_{c}^{2}\right)^{3}
\exp\left(-\frac{s}{T^{2}}\right)  \nonumber \\
&&+\frac{m_s \langle\bar{s}s\rangle}{384\pi^6}
\int dydz\,y(1-y-z)^2\left(s-\bar{m}_c^2\right)^2 (5s-3\bar{m}_c^2)
\exp\left(-\frac{s}{T^2}\right) \, ,
\end{eqnarray}

\begin{eqnarray}
\tilde{\rho}_4^0(s)&=&
-\frac{m_{c}^{2}\left\langle\tilde{\alpha} GG\right\rangle}{9216\pi^{6}}
\int dydz \frac{(1-y-z)^{4}}{y^{2}}	\left(1+\frac{z}{y}\right)		 \left(s-\bar{m}_{c}^{2}\right)\left(2s-\bar{m}_{c}^{2}\right)
\exp\left(-\frac{s}{T^{2}}\right) \nonumber \\
&&+\frac{\left\langle\tilde{\alpha} GG\right\rangle}{18432\pi^{6}}
\int dydz\frac{z(1-y-z)^{4}}{y^{2}}\left(s-\bar{m}_{c}^{2}\right)^{2}		 \left(5s-2\bar{m}_{c}^{2}\right)\exp\left(-\frac{s}{T^{2}}\right) \nonumber \\	
&&-\frac{\left\langle\tilde{\alpha} GG\right\rangle}{6144\pi^{6}}
\int dydz\, y(1-y-z)^{2}\left(s-\bar{m}_{c}^{2}\right)^{2}	 \left(5s-2\bar{m}_{c}^{2}\right)\exp\left(-\frac{s}{T^{2}}\right) \nonumber \\
&&+\frac{\left\langle\tilde{\alpha} GG\right\rangle}{9216\pi^{6}}
\int dydz\,(1-y-z)^{3}\left(s-\bar{m}_{c}^{2}\right)^{2}		 \left(5s-2\bar{m}_{c}^{2}\right)\exp\left(-\frac{s}{T^{2}}\right) \nonumber \\
&&- \frac{m_s m_c^3 \left\langle \tilde{\alpha} GG\right\rangle}{2304\pi^6}
\int dydz\, \frac{(1-y-z)^3}{y^3} \left(s-\bar{m}_c^2\right)
\exp\left(-\frac{s}{T^2}\right) \nonumber\\
&& + \frac{m_s m_c \left\langle \tilde{\alpha} GG\right\rangle}{1536\pi^6}
\int dydz\,\frac{(1-y-z)^3}{y^2} \left(s-\bar{m}_c^2\right)^2
\exp\left(-\frac{s}{T^2}\right) \nonumber\\
&& - \frac{7m_s m_c \left\langle \tilde{\alpha} GG\right\rangle}{4096\pi^6}
\int dydz\,(1-y-z)\left(s-\bar{m}_c^2\right)^2\exp\left(-\frac{s}{T^2}\right) \nonumber\\
&&+ \frac{m_s m_c \left\langle \tilde{\alpha} GG\right\rangle}{1024\pi^6}
\int dydz\,\frac{(1-y-z)^2}{y}\left(s-\bar{m}_c^2\right)^2
\exp\left(-\frac{s}{T^2}\right) \, ,			
\end{eqnarray}

\begin{eqnarray}
\tilde{\rho}_5^0(s)&=& \frac{3m_{c}\langle\bar{s}Gs\rangle}{512\pi^{6}}
 \int dydz\,(1-y-z)\left(s-\bar{m}_{c}^{2}\right)^{2}		 \exp\left(-\frac{s}{T^{2}}\right)  \nonumber \\
&&+\frac{m_s \langle\bar{s}Gs\rangle}{256\pi^6}\int dydz\,
y(1-y-z)(s-\bar{m}_c^2) (2s-\bar{m}_c^2)\exp\left(-\frac{s}{T^2}\right) \nonumber\\
&& - \frac{m_s \langle\bar{s}Gs\rangle}{256\pi^6}\int dydz\,
(1-y-z)^2(s-\bar{m}_c^2) (2s-\bar{m}_c^2)\exp\left(-\frac{s}{T^2}\right) \, ,	 \end{eqnarray}

\begin{eqnarray}
\tilde{\rho}_6^0(s)&=& \frac{\langle\bar{s}s\rangle^{2}}{24\pi^{4}}
\int dydz\, y(1-y-z)\left(s-\bar{m}_{c}^{2}\right)\left(2s-\bar{m}_{c}^{2}\right)
\exp\left(-\frac{s}{T^{2}}\right)  \nonumber \\
&&+\frac{5 m_s m_c \langle\bar{s}s\rangle^2}{24\pi^4}\int dydz\,
\left(s-\bar{m}_c^2\right)\exp\left(-\frac{s}{T^2}\right) \, ,
\end{eqnarray}

\begin{eqnarray}
\tilde{\rho}_7^0(s)&=&  \frac{m_{c}^{3}\langle\bar{s}s\rangle
\left\langle\tilde{\alpha} GG\right\rangle}{576\pi^{4}}
 \int dydz \frac{(1-y-z)^{2}}{y^{3}}\exp\left(-\frac{s}{T^{2}}\right)  \nonumber \\
&&-\frac{m_{c}\langle\bar{s}s\rangle\left\langle\tilde{\alpha} GG\right\rangle}{192\pi^{4}}
 \int dydz\frac{(1-y-z)^{2}}{y^{2}}	\left(s-\bar{m}_{c}^{2}\right)
\exp\left(-\frac{s}{T^{2}}\right) \nonumber \\		
&&+\frac{17m_{c}\langle\bar{s}s\rangle
\left\langle\tilde{\alpha} GG\right\rangle}{4608\pi^{4}}
\int dydz\left(s-\bar{m}_{c}^{2}\right)\exp\left(-\frac{s}{T^{2}}\right) \nonumber \\
&&-\frac{m_{c}\langle\bar{s}s\rangle\left\langle\tilde{\alpha} GG\right\rangle}{192\pi^{4}}
\int dydz\frac{(1-y-z)}{y}\left(s-\bar{m}_{c}^{2}\right)
\exp\left(-\frac{s}{T^{2}}\right) \nonumber \\	
&& -\frac{m_s m_c^2 \langle\bar{s}s\rangle \left\langle \tilde{\alpha} GG\right\rangle}{576\pi^4}\int dydz\,	\frac{(1-y-z)^2}{y^2} \left(1+\frac{z}{y}\right)			 \left[1+\frac{s}{2}\delta(s-\bar{m}_c^2)\right]
\exp\left(-\frac{s}{T^2}\right) \nonumber\\
&&+ \frac{ m_s \langle\bar{s}s\rangle \left\langle \tilde{\alpha} GG\right\rangle}{384\pi^4}	\int dydz\, \frac{z(1-y-z)^2}{y^2}
\left(3s-2\bar{m}_c^2\right)\exp\left(-\frac{s}{T^2}\right)\nonumber \\
&&-  \frac{7m_s \langle\bar{s}s\rangle \left\langle \tilde{\alpha} GG\right\rangle}{2304\pi^4}\int dydz\,y	\left(3s-2\bar{m}_c^2\right)
\exp\left(-\frac{s}{T^2}\right) \nonumber\\
&&+\frac{5 m_s \langle\bar{s}s\rangle \left\langle \tilde{\alpha} GG\right\rangle}{768\pi^4}\int dydz\,(1-y-z)\left(3s-2\bar{m}_c^2\right)
\exp\left(-\frac{s}{T^2}\right)\, ,
\end{eqnarray}

\begin{eqnarray}
\tilde{\rho}_8^0(s)&=&-\frac{5\langle\bar{s}s\rangle\langle\bar{s}Gs\rangle}{192\pi^{4}}
\int dydz\,y \left(3s-2\bar{m}_{c}^{2}\right)
\exp\left(-\frac{s}{T^{2}}\right) \nonumber \\	
&&+\frac{\langle\bar{s}s\rangle\langle\bar{s}Gs\rangle }{96\pi^{4}}
 \int dy dz (1-y-z)	 \left(3s-2\bar{m}_{c}^{2}\right)
\exp\left(-\frac{s}{T^{2}}\right) \nonumber \\
&& -\frac{121 m_s m_c \langle\bar{s}s\rangle\langle\bar{s}Gs\rangle}{1152\pi^4}
\int dy\,\exp\left(-\frac{s}{T^2}\right) \nonumber \\
&& + \frac{m_s m_c \langle\bar{s}s\rangle\langle\bar{s}Gs\rangle}{24\pi^4}
\int dydz\, \frac{1}{y}	\exp\left(-\frac{s}{T^2}\right)  \, ,
\end{eqnarray}

\begin{eqnarray}
\tilde{\rho}_9^0(s)&=&-\frac{m_{c}\langle\bar{s}s\rangle^{3}}{9\pi^{2}}
\int dy	\exp\left(-\frac{s}{T^{2}}\right)   \nonumber \\
&&+\frac{m_s \langle\bar{s}s\rangle^3}{18\pi^2}	\int dy\,
y\left[ 1+\frac{s}{2}\delta(s-\tilde{m}_c^2) \right]
\exp\left(-\frac{s}{T^2}\right) \, ,
\end{eqnarray}

\begin{eqnarray}
\tilde{\rho}_{10}^0(s)&=& \frac{\langle\bar{s}Gs\rangle^{2}}{192\pi^{4}}
 \int dy\, y\left[1+\frac{s}{2}\delta\left(s-\tilde{m}_{c}^{2}\right)		\right]	 \exp\left(-\frac{s}{T^{2}}\right)  \nonumber \\
&&-\frac{m_{c}^{2}\langle\bar{s}s\rangle^{2}\left\langle\tilde{\alpha} GG\right\rangle}{432\pi^{2}}
 \int dydz \frac{(1-y-z)}{y^{2}}\left(1+\frac{z}{y}\right)
\left(1+\frac{s}{T^{2}}\right)\delta\left(s-\bar{m}_{c}^{2}\right) \exp\left(-\frac{s}{T^{2}}\right) \nonumber \\
&&+\frac{\langle\bar{s}s\rangle^{2}\left\langle\tilde{\alpha} GG\right\rangle}{72\pi^{2}}
 \int dydz	\frac{z(1-y-z)}{y^{2}} \left[1+\frac{s}{2}\delta\left(s-\bar{m}_{c}^{2}\right)\right]
\exp\left(-\frac{s}{T^{2}}\right) \nonumber \\ 		
&&-\frac{\langle\bar{s}s\rangle^{2}\left\langle\tilde{\alpha} GG\right\rangle}{576\pi^{2}}
 \int dydz \left[1+\frac{s}{2}\delta\left(s-\bar{m}_{c}^{2}\right)\right]
\exp\left(-\frac{s}{T^{2}}\right) \nonumber \\
&&+\frac{\langle\bar{s}Gs\rangle^{2}}{384\pi^{4}} \int dy\,y \left[1+\frac{s}{2}\delta\left(s-\tilde{m}_{c}^{2}\right)\right]
\exp\left(-\frac{s}{T^{2}}\right) \nonumber \\
&&-\frac{5\langle\bar{s}Gs\rangle^{2}}{768\pi^{4}} \int dydz \left[1+\frac{s}{2}\delta\left(s-\bar{m}_{c}^{2}\right)\right]
\exp\left(-\frac{s}{T^{2}}\right) \nonumber \\	
&&+\frac{\langle\bar{s}s\rangle^{2}\left\langle\tilde{\alpha} GG\right\rangle}{216\pi^{2}}
\int dy\,y \left[1+\frac{s}{2}\delta\left(s-\tilde{m}_{c}^{2}\right)\right]
\exp\left(-\frac{s}{T^{2}}\right) \nonumber \\	
&&+\frac{31m_s m_c \langle\bar{s}Gs\rangle^2}{2304\pi^4}\int dy\,
\left(1+\frac{s}{T^2} \right)		 \delta(s-\tilde{m}_c^2)\exp\left(-\frac{s}{T^2}\right)\nonumber \\
&& - \frac{5 m_s m_c^3 \langle\bar{s}s\rangle^2 \left\langle\tilde{\alpha} GG\right\rangle}{216\pi^2 T^2}\int dydz\, \frac{1}{y^3}
\delta(s-\tilde{m}_c^2)\exp\left(-\frac{s}{T^2}\right) \nonumber \\
&& + \frac{5 m_s m_c \langle\bar{s}s\rangle^2 \left\langle \tilde{\alpha} GG\right\rangle}{72\pi^2}\int dydz\,\frac{1}{y^2}
\delta(s-\tilde{m}_c^2)\exp\left(-\frac{s}{T^2}\right) \nonumber\\
&& -\frac{ m_s m_c \langle\bar{s}s\rangle^2 \left\langle \tilde{\alpha} GG\right\rangle}{288\pi^2}\int dy\,\frac{1}{y}
\delta(s-\tilde{m}_c^2)\exp\left(-\frac{s}{T^2}\right) \nonumber\\
&& - \frac{m_s m_c \langle\bar{s}Gs\rangle^2}{96\pi^4}\int dy\,
\frac{1}{y}	\delta(s-\tilde{m}_c^2)\exp\left(-\frac{s}{T^2}\right) \nonumber\\
&& +\frac{11 m_s m_c \langle\bar{s}s\rangle^2 \left\langle \tilde{\alpha} GG\right\rangle}{864\pi^2}\int dy\,\left( 1+\frac{s}{T^2} \right)
\delta(s-\tilde{m}_c^2)\exp\left(-\frac{s}{T^2}\right) \, ,
\end{eqnarray}

\begin{eqnarray}
\tilde{\rho}_{11}^0(s)&=&\frac{m_{c}\langle\bar{s}s\rangle^{2}\langle\bar{s}Gs\rangle
}{12\pi^{2}} \int dy\left(1+\frac{s}{T^{2}}	\right)	 \delta\left(s-\tilde{m}_{c}^{2}\right) \exp\left(-\frac{s}{T^{2}}\right)   \nonumber \\
&&-\frac{m_{c}\langle\bar{s}s\rangle^{2}\langle\bar{s}Gs\rangle}{36\pi^{2}}
 \int dy	\frac{1}{y}	\delta\left(s-\tilde{m}_{c}^{2}\right) \exp\left(-\frac{s}{T^{2}}\right) \nonumber \\	
&&-\frac{m_s \langle\bar{s}s\rangle^2\langle\bar{s}Gs\rangle}{27\pi^2}
\int dy\,y\left( 1+\frac{s}{T^2}+\frac{s^2}{2T^4} \right)
\delta(s-\tilde{m}_c^2)\exp\left(-\frac{s}{T^2}\right) \nonumber\\
&& +\frac{m_s \langle\bar{s}s\rangle^2\langle\bar{s}Gs\rangle}{144\pi^2 }
\int dy\,\left( 1+\frac{s}{T^2} \right)
\delta(s-\tilde{m}_c^2)\exp\left(-\frac{s}{T^2}\right)  \, ,
\end{eqnarray}

\begin{eqnarray}
\tilde{\rho}_{13}^0(s)&=&-\frac{m_{c}\langle\bar{s}s\rangle\langle\bar{s}Gs\rangle^{2}
}{48\pi^{2}T^{6}} \int dy\,s^2 \delta\left(s-\tilde{m}_{c}^{2}\right) \exp\left(-\frac{s}{T^{2}}\right) \nonumber \\
&&+\frac{m_{c}^{3}\langle\bar{s}s\rangle^{3}
\left\langle\tilde{\alpha} GG\right\rangle}{81T^{4}} \int dy\frac{1}{y^{3}}
\delta\left(s-\tilde{m}_{c}^{2}\right)\exp\left(-\frac{s}{T^{2}}\right) \nonumber \\
&&-\frac{m_{c}\langle\bar{s}s\rangle^{3}
\left\langle\tilde{\alpha} GG\right\rangle}{27T^{2}}
 \int dy\frac{1}{y^{2}} \delta\left(s-\tilde{m}_{c}^{2}\right) \exp\left(-\frac{s}{T^{2}}\right) \nonumber \\
&&+\frac{m_{c}\langle\bar{s}s\rangle\langle\bar{s}Gs\rangle^{2}}{72\pi^{2}T^{4}}
 \int dy\frac{1}{y}	\,s\, \delta\left(s-\tilde{m}_{c}^{2}\right) \exp\left(-\frac{s}{T^{2}}\right) \nonumber \\
&&-\frac{m_{c}\langle\bar{s}s\rangle^{3}
\left\langle\tilde{\alpha} GG\right\rangle}{108T^{6}}
\int dy\, s^2\, \delta\left(s-\tilde{m}_{c}^{2}\right) \exp\left(-\frac{s}{T^{2}}\right) \nonumber \\		
&&+ \frac{7 m_s \langle\bar{s}s\rangle\langle\bar{s}Gs\rangle^2}{1728\pi^2 T^8}
\int dy\,y\,s^3\, \delta(s-\tilde{m}_c^2)\exp\left(-\frac{s}{T^2}\right)\nonumber  \\
&& + \frac{m_s m_c^2 \langle\bar{s}s\rangle^3 \left\langle \tilde{\alpha} GG\right\rangle}{648T^4}\int dy\,\frac{1}{y^3}\left( 1-\frac{s}{T^2} \right)
\delta(s-\tilde{m}_c^2)\exp\left(-\frac{s}{T^2}\right) \nonumber\\
&& + \frac{m_s \langle\bar{s}s\rangle^3 \left\langle \tilde{\alpha} GG\right\rangle}{216T^4}\int dy\, \frac{(1-y)}{y^2}\, s\,
\delta(s-\tilde{m}_c^2)\exp\left(-\frac{s}{T^2}\right)\nonumber \\
&&- \frac{5 m_s \langle\bar{s}s\rangle\langle\bar{s}Gs\rangle^2}{1728\pi^2 T^6}
\int dy\,s^2\, \delta(s-\tilde{m}_c^2)\exp\left(-\frac{s}{T^2}\right)\nonumber \\
&&+ \frac{m_s \langle\bar{s}s\rangle^3 \left\langle \tilde{\alpha} GG\right\rangle}{648T^8}\int dy\,y\, s^3\, 			
\delta(s-\tilde{m}_c^2)\exp\left(-\frac{s}{T^2}\right)\, ,
\end{eqnarray}
 $\langle\bar{s} Gs\rangle=\langle\bar{s}g_s\sigma Gs\rangle$, $\tilde{\alpha}_s=\frac{\alpha_s}{\pi}$,
$\int dydz=\int_{y_{i}}^{y_{f}} dy \int_{z_{i}}^{1-y} dz$,
$dy=\int_{y_{i}}^{y_{f}} dy $, $y_{f}=\frac{1+\sqrt{1-4m_c^2/s}}{2}$, $y_{i}=\frac{1-\sqrt{1-4m_c^2/s}}{2}$, $z_{i}=\frac{ym_c^2}{y s -m_c^2}$, $\bar{m}_c^2=\frac{(y+z)m_c^2}{yz}$,
$ \tilde{m}_c^2=\frac{m_c^2}{y(1-y)}$, and $\int_{y_i}^{y_f}dy \to \int_{0}^{1}dy$, $\int_{z_i}^{1-y}dz \to \int_{0}^{1-y}dz$ when the $\delta$ functions $\delta\left(s-\bar{m}_c^2\right)$ and $\delta\left(s-\tilde{m}_c^2\right)$ appear.

\section*{Acknowledgements}
This  work is supported by National Natural Science Foundation, Grant Number  12575083.


\begin{thebibliography}{99}

\bibitem{X3872-2003} S. K. Choi   et al, Phys. Rev. Lett. {\bf 91} (2003) 262001.

\bibitem{LHCb-4380} R. Aaij  et al, Phys. Rev. Lett. {\bf 115} (2015) 072001.

\bibitem{LHCb-Pc4312} R. Aaij et al, Phys. Rev. Lett. {\bf 122} (2019) 222001.

\bibitem{LHCb-Pcs4459-2012} R. Aaij  et al, Sci. Bull. {\bf 66} (2021) 1278.

\bibitem{LHCb-Pc4337}  R. Aaij  et al,  Phys. Rev. Lett. {\bf 128} (2022) 062001.

\bibitem{LHCb-Pcs4338}  R. Aaij et al, Phys. Rev. Lett. {\bf 131} (2023)  031901.

\bibitem{Belle-Pcs4338-Pcs4459} I. Adachi et al,  Phys. Rev. Lett. {\bf 135} (2025) 041901.

\bibitem{di-di-anti-penta-1} L. Maiani, A. D. Polosa and V. Riquer,  Phys. Lett. {\bf B749} (2015) 289.

\bibitem{di-di-anti-penta-2} G. N. Li, M. He and X. G. He,  JHEP {\bf 1512} (2015) 128.

\bibitem{Wang1508-EPJC} Z. G. Wang, Eur. Phys. J. {\bf C76} (2016) 70.

\bibitem{WangHuang-EPJC-1508-12} Z. G. Wang  and T. Huang, Eur. Phys. J. {\bf C76} (2016)  43.

\bibitem{WangZG-EPJC-1509-12} Z. G. Wang, Eur. Phys. J. {\bf C76} (2016)  142.

\bibitem{WangZG-NPB-1512-32} Z. G. Wang, Nucl. Phys. {\bf B913} (2016) 163.


\bibitem{Pc4312-penta-3} R. Zhu, X. Liu, H. Huang and C. F. Qiao,  Phys. Lett. {\bf B797} (2019) 134869.

\bibitem{Pc4312-penta-1} A. Ali and A. Y. Parkhomenko, Phys. Lett. {\bf B793} (2019) 365.


\bibitem{WZG-penta-IJMPA} Z. G. Wang, Int. J. Mod. Phys. {\bf A35} (2020)  2050003.

\bibitem{WangZG-Pcs4459-333} Z. G. Wang, Int. J. Mod. Phys. {\bf A36} (2021) 2150071.


\bibitem{WangZG-Pc12-JpsiLambda}  Z. G. Wang and Q. Xin, Eur. Phys. J. {\bf C86} (2026)  472.

\bibitem{WangZG-Pc12-Jpsip}  Z. G. Wang, Eur. Phys. J. {\bf C86} (2026) 209.

\bibitem{WangZG-Pc12-JpsiXi}  Z. G. Wang and Y. Liu, Nucl. Phys. {\bf B1027} (2026) 117456.

\bibitem{WangZG-Pc12-JpsiSgm}  Z. G. Wang and Y. Liu, Front. Phys. {\bf 22} (2027) 016201.

\bibitem{WangZG-Pc12-JpsiXi-10} Z. G. Wang, arXiv: 2606.28095 [hep-ph].


\bibitem{di-tri-penta-1} R. F. Lebed, Phys. Lett. {\bf B749} (2015) 454.

\bibitem{di-tri-penta-2} R. Zhu and C. F. Qiao, Phys. Lett. {\bf B756} (2016) 259.



\bibitem{mole-penta-2}  R. Chen, X. Liu, X. Q. Li and S. L. Zhu, Phys. Rev. Lett. {\bf 115} (2015) 132002.


\bibitem{mole-penta-1}  F. K. Guo, U. G. Meissner, W. Wang  and Z. Yang, Phys. Rev. {\bf D92} (2015) 071502.

\bibitem{mole-penta-3}  H. X. Chen, W. Chen, X. Liu, T. G. Steele and S. L. Zhu, Phys. Rev. Lett. {\bf 115} (2015)  172001.

\bibitem{mole-penta-10}   K. Azizi, Y. Sarac and H. Sundu, Phys. Lett. {\bf B782} (2018) 694.

\bibitem{mole-penta-11}  Z. G. Wang,   Int. J. Mod. Phys. {\bf A34} (2019)  1950097.

\bibitem{mole-penta-6}  T. J. Burns, Eur. Phys. J. {\bf A51} (2015)  152.

\bibitem{mole-penta-5}  J. He, Phys. Lett. {\bf B753} (2016) 547.

\bibitem{Pc4312-mole-penta-2}  C. W. Xiao, J. Nieves and E. Oset, Phys. Rev.
{\bf D100} (2019)  014021.

\bibitem{Pc4312-mole-penta-1}  M. Z. Liu et al, Phys. Rev. Lett. {\bf 122} (2019) 242001.

\bibitem{Pc4312-mole-penta-WXW-SCPMA}  X. W. Wang, Z. G. Wang, G. L. Yu and Q. Xin, Sci. China-Phys. Mech. Astron. {\bf 65} (2022) 291011.

\bibitem{Pc4312-mole-penta-WXW-IJMPA} X. W. Wang and Z. G. Wang, Int. J. Mod. Phys. {\bf A37} (2022) 2250189.

\bibitem{Pcs4338-mole-XWWang} X. W. Wang and Z. G. Wang, Chin. Phys. {\bf C47} (2023)  013109.


\bibitem{Pc4312-mole-penta-7}   Z. H. Guo and J. A. Oller, Phys. Lett. {\bf B793} (2019) 144.


\bibitem{Pcs4338-mole-FKGuo} X. K. Dong, F. K. Guo and B. S. Zou, Commun. Theor. Phys. {\bf 73} (2021)  125201.

\bibitem{Pcs4459-mole-FLWang}   F. L. Wang and X. Liu, Phys. Lett. {\bf B835} (2022) 137583.

\bibitem{Pcs4459-mole-CWXiao}  C. W. Xiao, J. J. Wu and B. S. Zou, Phys. Rev. {\bf D103} (2021)  054016.

\bibitem{Coupled-Pc4337-MJYan-EPJC-2022}  M. J. Yan, F. Z. Peng, M. S. Sanchez and M. P. Valderrama,  Eur. Phys. J. {\bf C82} (2022)  57.

 \bibitem{Pcs4338-mole-LMeng} L. Meng, B. Wang and S. L. Zhu, Phys. Rev. {\bf D107} (2023)  014005.


\bibitem{ATS-LiuXH-2016}  X. H. Liu, Q. Wang and Q. Zhao,  Phys. Lett. {\bf B757} (2016) 231.

\bibitem{ATS-Bayar-GuoFK-2016} M. Bayar, F. Aceti, F. K. Guo and E. Oset,  Phys. Rev. {\bf D94} (2016)  074039.

\bibitem{ATS-LiuXH-2020} F. K. Guo, X. H. Liu and S. Sakai, Prog. Part. Nucl. Phys. {\bf 112} (2020) 103757.

\bibitem{WangZG-Review} Z. G. Wang,  Front. Phys. {\bf 21} (2026) 016300.


\bibitem{Zong-penta-sss-PLB-19} Q. Meng, E. Hiyama, K. U. Can, P. Gubler, M. Oka, A. Hosaka and H. S. Zong, Phys. Lett. {\bf B798} (2019) 135028.

\bibitem{Li-penta-sss-PRD-23}
S. Y. Li, Y. R. Liu, Z. L. Man, Z. G. Si and  J. Wu, Phys. Rev. {\bf D108} (2023)  056015.

\bibitem{WangFL-mole-sss-PRD-21} F. L. Wang, X. D. Yang, R. Chen and X. Liu,
Phys. Rev. {\bf D103} (2021)  054025.


\bibitem{Clymton-mole-sss-PRD-25}
    S. Clymton, H. C. Kim and T. Mart, Phys. Rev. {\bf D112} (2025)  094024.

\bibitem{Oset-mole-sss-PRD-24}
  L. Roca, J. Song and Oset, Phys. Rev. {\bf D109} (2024)  094005.

\bibitem{Azizi-penta-sss-PRD-23} K. Azizi, Y. Sarac and H. Sundu, Phys. Rev.
{\bf D107} (2023)  014023.

\bibitem{WangZG-landau-PRD} Z.~G.~Wang, Phys. Rev. {\bf D101} (2020)  074011.


\bibitem{SVZ79-1}  M. A. Shifman, A. I. Vainshtein and V. I. Zakharov, Nucl. Phys. {\bf B147} (1979) 385.

\bibitem{SVZ79-2}  M. A. Shifman, A. I. Vainshtein and V. I. Zakharov, Nucl. Phys. {\bf B147} (1979) 448.


\bibitem{PRT85} L. J. Reinders, H. Rubinstein and S. Yazaki, Phys. Rept. {\bf 127} (1985) 1.


\bibitem{WangHuang3900}  Z. G. Wang and T. Huang,  Phys. Rev. {\bf D89} (2014)  054019.

\bibitem{Wang-tetra-NPB-HCss} Z. G. Wang, Nucl. Phys. {\bf B1007} (2024) 116661.

\bibitem{WangZG-IJMPA-2021}  Z. G. Wang,  Int. J. Mod. Phys. {\bf A36} (2021) 2150107.

\bibitem{ColangeloReview} P. Colangelo and A. Khodjamirian, hep-ph/0010175.

\bibitem{PDG}  F. Takahashi et al,  Int. J. Mod. Phys. {\bf A41} (2026)  2630011.

\bibitem{Narison-mix} S. Narison and R. Tarrach, Phys. Lett. {\bf 125 B} (1983) 217.

\bibitem{Wang-tetra-formula}  Z. G. Wang, Eur. Phys. J. {\bf C74} (2014)  2874.

\bibitem{WangZG-mole-formula-1} Z. G. Wang and T. Huang, Eur. Phys. J. {\bf C74} (2014)  2891.

\bibitem{WangZG-mole-formula-2}  Z. G. Wang,  Eur. Phys. J. {\bf C74} (2014)  2963.

\bibitem{WangZG-IJMPA-3-scheme} Z. G. Wang, Int. J. Mod. Phys. {\bf A34} (2019)  1950097.

\bibitem{Pascual-1984} P. Pascual and R. Tarrach, ``QCD: Renormalization for the practitioner", Springer Berlin Heidelberg (1984).

\bibitem{WangXW-penta-mole-sss} X. W. Wang and Z. G. Wang, in preparation.

\end{thebibliography}
\end{document}